\documentclass[fleqn,usenatbib]{mnras}

\usepackage{newtxtext,newtxmath}

\usepackage[T1]{fontenc}
\usepackage{ae,aecompl}

\makeatletter

\usepackage{graphicx}	
\usepackage{amsmath}	
\usepackage[normalem]{ulem}
\usepackage{soul}

\newcommand{\gaia}{$Gaia\,$}

\title[Undiscovered Outer Globular Clusters]{The Likelihood of Undiscovered Globular Clusters in the Outskirts of the Milky Way}
\author[Webb \& Carlberg]{Jeremy J. Webb$^1$ \thanks{E-mail: webb@astro.utoronto.ca (JW)} \& Raymond G. Carlberg$^1$ \\
$^1$Department of Astronomy and Astrophysics, University of Toronto, 50 St. George Street, Toronto, ON, M5S 3H4, Canada \\
}

\date{Accepted XXX. Received YYY; in original form ZZZ}

\pubyear{2020}

\begin{document}
\label{firstpage}
\pagerange{\pageref{firstpage}--\pageref{lastpage}}
\maketitle

\begin{abstract}
The currently known Galactic globular cluster population extends out to a maximum galactocentric distance of $\sim$ 145 kpc, with the peculiarity that the outermost clusters predominantly have an inward velocity. Orbit averaging finds that this configuration occurs by chance about 6\% of the time, suggesting that several globular clusters with positive radial velocities remain undiscovered. We evaluate the expected number of undiscovered clusters at large distances under the assumption that the cluster population has a smooth radial distribution and is in equilibrium within the Milky Way's virial radius. By comparing the present day kinematic properties of outer clusters to random orbital configurations of the Galactic globular cluster system through orbit averaging, we estimate a likelihood of $73\%$ of there being at least one undiscovered globular cluster within the Milky Way. This estimate assumes the current population is complete out to 50 kpc, and increases to $91\%$ if the population is complete out to 150 kpc. The likelihood of there being two undiscovered clusters is between  $60\%$ and $70\%$, with the likelihood of there being three undiscovered clusters being on the order of $50\%$. The most likely scenario is that the undiscovered clusters are moving outwards, which results in the outer cluster population being consistent with an equilibrium state. Searches for distant and possibly quite low concentration and very low metallicity globular clusters will be enabled with upcoming deep imaging surveys. 
\end{abstract}


\section{Introduction} \label{s_intro}

The first globular cluster discovered is believed to be M22 by amateur astronomer Abraham Ihle \citep{Lynn1886}. However at the time it was identified as a nebula because individual cluster stars could not be resolved \citep{Halley1716}. Over 100 years later William Herschel was able to resolve individual clusters and introduced the term globular cluster for a large sky survey that contained 70 clusters (36 of which were newly discovered) \citep{Herschel1789}. \citet{Herschel1789} wrote, "we come to know that there are globular clusters of stars nearly equal in size, which are scattered evenly at equal distances from the middle, but with an increasing accumulation towards the center.". These early studies proved extremely valuable to the study of both globular clusters and galaxies. In the early 1900's, through a series of paper, Howard Shapley made use of the known globular clusters in the Milky Way to study its shape and structure \citep[e.g][]{Shapley18a,Shapley18b}. The  \citet{Plaskett1939} Milky Way review paper included his famous Plate XIV, and noted that the kinematics suggested a significant fraction of the Milky Way's mass is in the form of "diffuse matter".

In the 230 years since the work of \citet{Herschel1789}, the number of discovered Galactic globular clusters has more than doubled as the imaging depth and image quality of all-sky surveys have allowed for more and more distant clusters to be discovered. The 1960's and 1970's new photographic sky surveys allowed the discovery of the distant and faint Palomar clusters \citep{Arp60} and AM-1 \citep{Madore79}, while more recently the Sloan Digital Sky Survey led to the discovery of  several faint globular clusters within 60 kpc of the Galactic centre. The  \citet{Harris96,Harris10} catalog of globular cluster parameters lists 157 known globular clusters in the Milky Way.

A large number of confirmed and candidate globular clusters have been discovered since the \citet{Harris96,Harris10} catalogue was released \citep[e.g][]{Ryu18,Gran2019}. The majority of these candidate clusters are close to the Galactic plane with significant dust extinction and have only recently been discovered with the help of IR surveys. The VISTA Variables in the Vía Láctea survey \citep{Minniti10} helped locate the majority of these candidate clusters in the direction of the Galactic bulge. \citet{Ryu18} also found two additional clusters close to the Galactic plane in the UKIRT Infrared Deep Sky Survey Galactic Plane Survey (Lawrence et al. 2007; Lucas et al. 2008). 

\citet{Ryu18} does, however, list $\sim$ 25 cluster candidates that are not associated with the Galatic plane. Several of these clusters, namely Segue 3 \citep{Belokurov2010}, Mu{\~n}oz 1 \citep{Munoz2012}, Balbinot 1 \citep{Balbinot2013}, Kim 1 \citep{Kim2015}, Kim 2 \citep{Kim2015b}, Eridanus III \citep{Bechtol2015_eri11}, DES 1 \citep{Luque2016_des1}, and Kim 3 \citep{Kim2016} are considered to be ultra-faint objects and its not clear whether or not they are globular clusters in the process of dissolution or ultra-faint dwarf galaxies \citep{Contenta17}. Using a globular cluster population synthesis model, \citet{Contenta17} estimates there to be $3^{+7.3}_{-1.6}$ faint star clusters beyond 20 kpc, which is consistent with the candidates listed above as being globular clusters. However, additional observations are required to confirm these ultra-faint candidates are indeed globular clusters.

Globular clusters with large heliocentric distances are difficult to discover because they are faint and usually fairly extended \citep{Laevens2014}. They are, however, relatively straightforward confirm as globular clusters once found since they are likely to be halo clusters at high galactic latitudes with little extinction. The most distant globular cluster, Crater (also known as Laevens 1), was discovered in 2014 with the Pan-STARRS1 survey \citep{Laevens2014, Chambers16}, and separately with the ESO VST ATLAS Survey \citep{Belokurov14}. Its discovery extends the Galactic globular cluster population out to a distance of $\sim 145$ kpc. The globular cluster Laevens 3 was also discovered with Pan-STARRS1 \citep{Laevens2015}, orbiting with a galactocentric distance of $\sim$ 58 kpc. Ultra-faint candidate clusters DES 1 \citep{Luque2016_des1}, Eridanus III \citep{Conn2018}, Kim 2 \citep{Kim2015b} both have galactocentric distances greater than 85 kpc, and may prove to also be members of the outer Galactic globular cluster population.

With more telescopes and planned surveys coming online in the near future (Vera C. Rubin Observatory, EUCLID, WFIRST), along with continual data releases from existing surveys like SDSS and \gaia, it is worth considering the likelihood of finding additional globular clusters in the Milky Way. Given how strongly globular clusters are linked to the formation and evolution of their host galaxy \citep{Baumgardt03, Kruijssen09}, and their recent use in constraining the Milky Way's merger history \citep{Myeong18, Massari19, Kruijssen20}, discovering new globular clusters will help solidify our understanding of our own Galaxy and its satellite population.

The incompleteness of the inner Galactic globular cluster population can be estimated given that significantly more clusters have been observed near the Galactic plane in the close-half of the Galaxy than in the far-half \citet{Ryu18}. This discrepancy continues to drive the search for undiscovered clusters near the Galactic plane. However estimating the population's completeness at large galactocentric distances is less straightforward. \citet{Youakim20} recently used the lack of metal poor globular clusters ([Fe/H] < -2.5) to suggest that either an additional 5.4 clusters may exist in the outer Galactic halo or the globular cluster metallicity distribution function has a metallicity floor. The recent discovery an extremely metal poor globular cluster in M31 with [Fe/H]=-2.91 \citep{Larsen2020} appears to support the idea that several undetected metal poor clusters exist in the Milky Way's outer halo. An alternative approach, only now possible with the release of \gaia DR2 \citep{Gaia16a,Gaia18}, is to make use of the orbital properties of outer Galactic globular clusters.

The purpose of this study is to use the spatial and kinematic properties of the outer Galactic globular cluster system to determine the likelihood that one or more distant clusters remain undiscovered. We specifically use the orbital properties of the Galactic globular cluster population to estimate the probability of any remaining undiscovered globular clusters in the outskirts of the Milky Way. One indicator of missing clusters is whether there are equal numbers of clusters with positive and negative radial velocities, which should be statistically equal for an outer population that is in equilibrium and complete. 

A second, but more complex, test for how complete the outer cluster population is to examine the properties of the orbit averaged distribution of several outer cluster properties. The present day distribution of cluster properties, such as the galactocentric radii and velocities, should not be a rare configuration.  Large  fluctuations with orbital phase of the observed population's radial and total velocity distributions suggests that several outer clusters are awaiting discovery. 

The outline of this study is as follows. In Section \ref{s_method} we summarize the properties of the known Galactic globular cluster population, namely its radial, velocity, and orbital distributions,  highlighting features that suggest the outer population is incomplete. In Section \ref{s_results} we consider how the distribution of cluster radii, velocities, and orbits vary as the cluster orbits are integrated in a static potential. Differences between the current and random future distributions of cluster properties will indicate the outer population is incomplete. We then estimate the likelihood of their being undiscovered Galactic globular clusters with large galactocentric distances by comparing the present day properties of outer globular clusters to the most probable distribution of outer cluster properties. The detectability and implications of finding undiscovered outer clusters, as well as a summary of our results, are discussed in Section \ref{s_conclusion}. 

\section{The Galactic Globular Cluster Population} \label{s_method}

The recent \citet{Baumgardt19} catalogue contains 159 clusters with their kinematic data \citep{Gaia16a,Gaia2018}. Hence it is possible to calculate the variations with orbital phase of the radial, velocity, and orbital distributions of Galactic clusters, and their averages, in a static potential. The catalogue contains 7 clusters that have been discovered since the most recent edition of the \citet{Harris96, Harris10} catalogue (Crater, BH 140, FSR 1716, FSR 1758, VVVCL001, Mercer 5,Laevens 3), but omits five clusters ((BH 176, GLIMPSE 1, GLIMPSE 2 Koposov 1, and Koposov 2). 

The proper motions of BH 176, whose classification as a globular cluster is debatable \citep{Davoust2011}, were measured by \citet{Vasiliev19} and are added to our dataset. \citet{Vasiliev19} notes that GLIMPSE 1 and GLIMPSE 2 are too heavily extincted to have their kinematics measured by \gaia. However since the focus of this study is the completeness of the outer Galactic cluster population, the exclusion of GLIMPSE 1 and GLIMPSE 2 from our dataset is inconsequential as they are inner clusters ($r<5$ kpc) in the plane of the disk ($\mid z \mid < 0.1$ kpc). The exclusion of Koposov 1 and Koposov 2, however, present a problem when studying the completeness of the halo cluster population as they have galactocentric distances of 49.3 kpc and 41.9 kpc respectively \citep{Harris10}. Unfortunately they are so distant that 
few cluster red giant stars are within the observational reach of \gaia. Therefore we exclude both clusters from our initial analysis, but consider how they may impact our results in Section \ref{s_koposov}. We also do not include any globular cluster candidates in our primary sample, but discuss their impact on our results in Section \ref{s_gccs}. Hence our primary sample comprises the 159 clusters in \citet{Baumgardt19}, along with BH 176 from \citet{Harris10} and \citet{Vasiliev19}.

As discussed in Section \ref{s_intro}, the Galactic globular cluster population is incomplete near the Galactic plane due to Galactic dust extinction. For the purposes of this study, we therefore further remove clusters with low Galactic latitudes, $b$, from our Galactic globular cluster dataset as the incompleteness would affect our estimates of the population's spatial and kinematic distributions. Such a restriction will not limit our analysis as the macroscopic properties of a spherically symmetric system will have no dependence on $b$. We exclude all clusters with $|b| < 10$, leaving a total of 86 clusters. We note that the exclusion of low latitude clusters ends up removing BH 176 from our final dataset, meaning the debate regarding its classification is of no consequence to this study. The final dataset does, however, include NGC 5139 (Omega Cen) which has been suggested to be the leftover core of a dwarf galaxy. Hence its inclusion in a dataset that is meant to be in equilibrium and complete may be erroneous. Therefore, in Section \ref{s_latcomplete} we address how an even more conservative cut affects our results. We assume that the distribution of outer halo clusters is spherically symmetric so that any cluster that orbits into the low latitude region has a statistical twin that orbits out of the low latitude region.

\begin{figure}
    \includegraphics[width=0.48\textwidth]{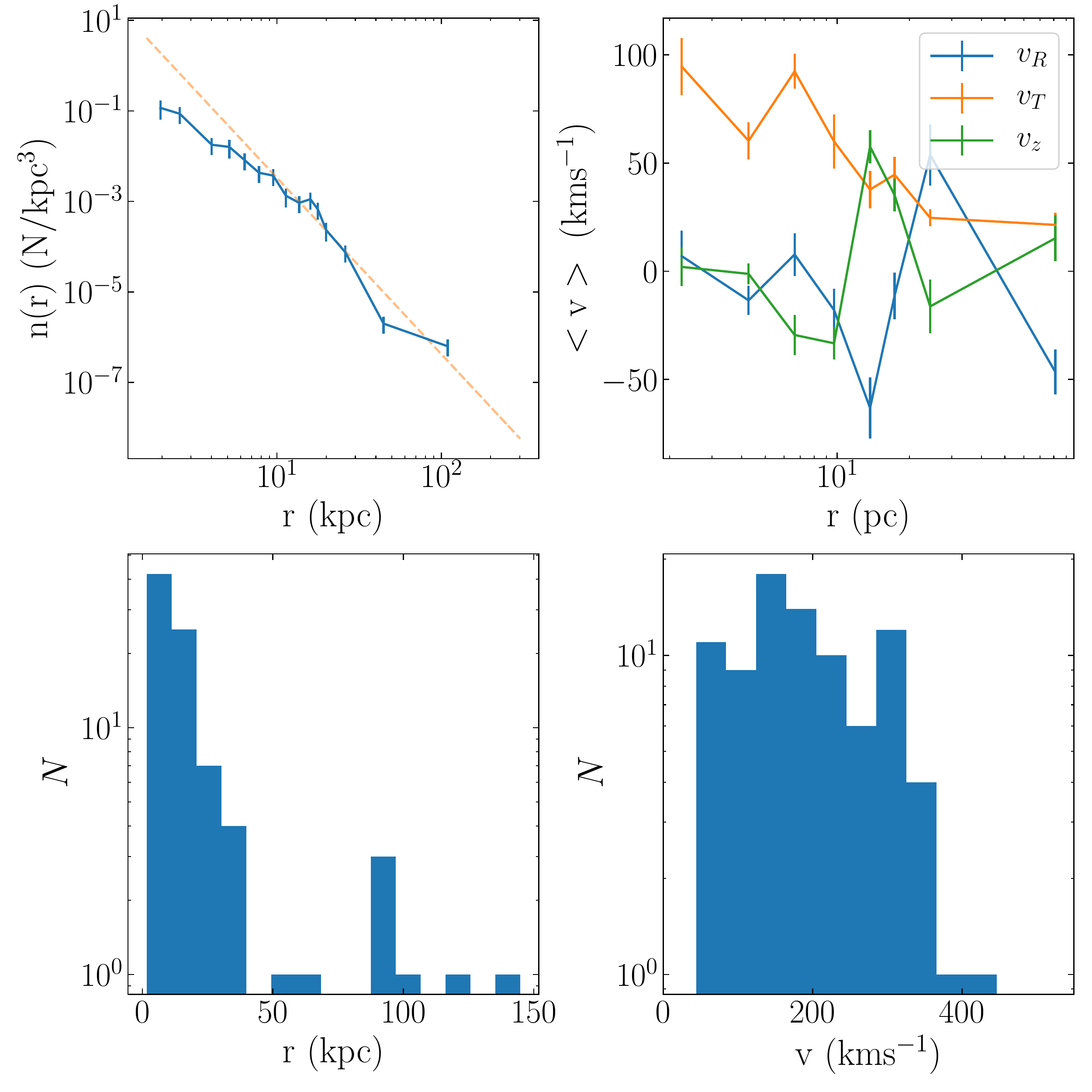}
    \caption{Number density (top left panel), radius distribution (bottom left panel), mean velocity profiles in cylindrical coordinate (top right panel) and total velocity distribution (bottom right panel) of the 86 Galactic globular clusters with $b>10^{\circ}$. In the top left panel, the number density profile has been fit with a power-law beyond 10 kpc. In the top-left panel error bars represent $1/\sqrt(N)$ error, while in the top-right panel error bars represent the standard error in the mean.}
    \label{fig:mwgc}
\end{figure}

The present day spatial and kinematic properties of our Galactic cluster dataset are illustrated in Figure \ref{fig:mwgc}, which specifically shows the population's number density profile, galactocentric radius distribution, mean velocity profiles in cylindrical coordinates, and total velocity distribution. The number density profile is determined by radially binning the clusters into 15 bins with an equal number of clusters, where the bin's location is set equal to the mean $r$ of all stars in the bin and errorbars illustrate $1/\sqrt(N)$ error. As indicated by the linear fit to the profile beyond 10 kpc, the outer number density profile follows a power-law of the form $n(r) \propto r ^{-3.9 \pm 0.4}$. The mean velocity profiles are determined by radially binning the clusters into 8 bins of equal size, where the error bars represent the error in the mean.

There are three specific features in Figure \ref{fig:mwgc} that raise questions as to whether the outer cluster population is incomplete:
\begin{itemize}
\item The distribution of galactocentric radii drops sharply near 50 kpc (bottom-left panel),
\item The mean cylindrical radial velocity $v_R$ profile, which should be centered around zero for a system that is in equilibrium and complete, drops to negative values at large $r$ (top-right panel), and,
\item The total velocity distribution is lacking in clusters with low velocities, which is surprising given how clusters spend the majority of their time at or near apocentre where $v \sim 0$ km/s (bottom-right panel). 
\end{itemize}

The mean cylindrical radial velocity,$v_R$, of the outermost 11 globular clusters is -46.5 km/s $\pm$ 10 km/s, statistically offset from $\langle v_R \rangle$ = 0. On the other hand, the velocity perpendicular to the plane, $v_z$, is consistent with zero as expected. Of these 11, 7 are moving inwards and only 4 are moving outwards. In spherical coordinates, the difference between the number of outer clusters with $v_r > 0$ and the number of outer clusters $v_r < 0$ remains the same, however the outer mean $\langle v_r \rangle$ is consistent with zero ($\langle v_{r} \rangle = -7 \pm 11$ km/s). At large galactoctocentric distance, it is the mean azimuthal velocity $\langle v_{\theta} \rangle$ that becomes inconsistent with zero ($\langle v_{\theta} \rangle = 40 \pm 10$ km/s), which is unexpected for a system in equilibrium.

\begin{figure}
    \includegraphics[width=0.48\textwidth]{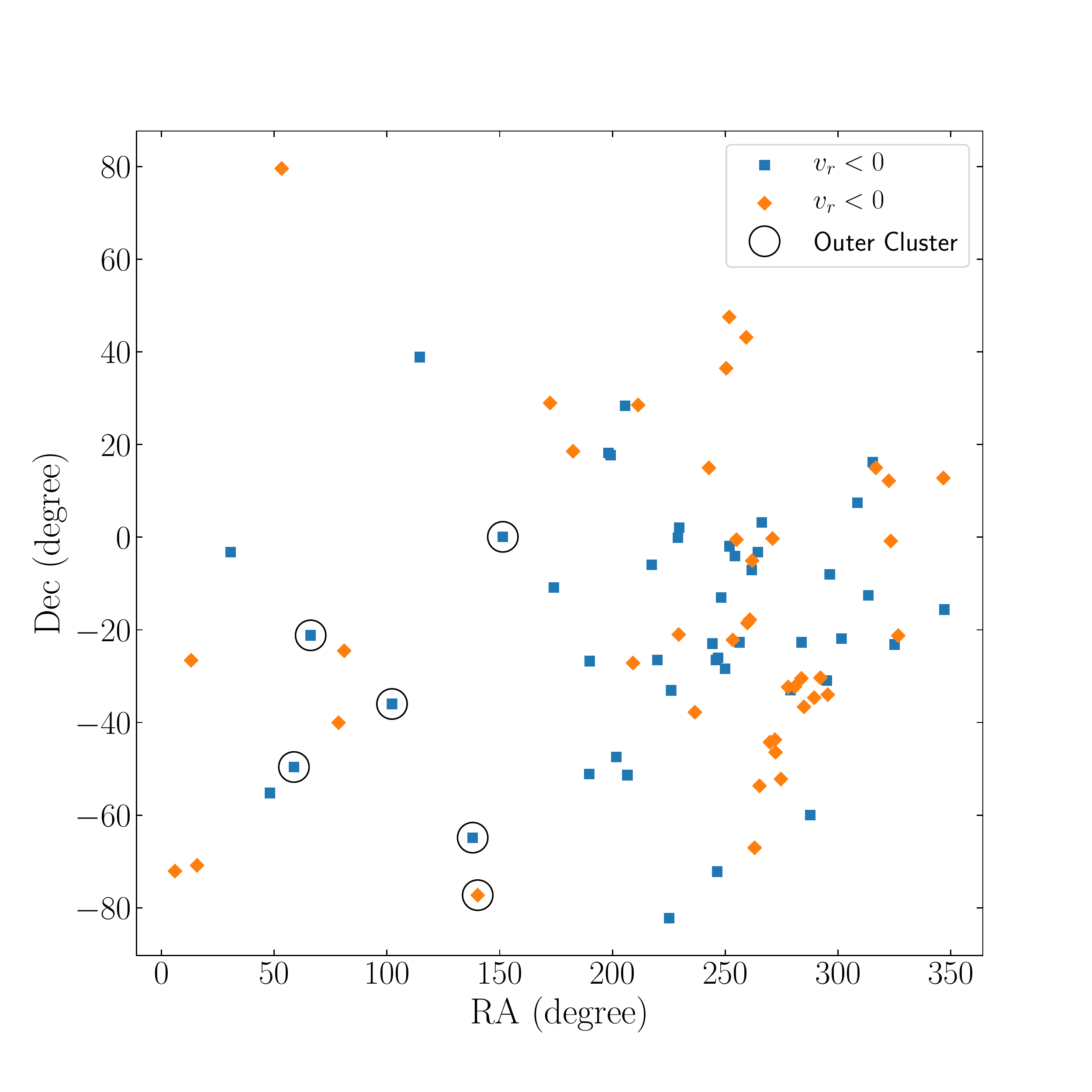}
    \caption{R.A. and Dec of the globular cluster dataset. Blue squares represent clusters that are moving inwards and orange diamonds represent clusters that are moving outwards. The outer 6 clusters have been circled.}
    \label{fig:radec}
\end{figure}

The difference between the number of clusters moving inwards and the number of globular clusters moving outwards (which we will refer to as $\Delta {N_{v_R}}$) reaches a maximum of 5 when only the outer 6 clusters are considered. NGC 2298, Eridanus, NGC 2808, AM1, and Pal3 are all have $v_R < 0$, while E3 (the third most distance globular cluster) has $v_R > 0$. The mean $\langle v_R \rangle$ of the 6 outermost clusters is -83.9 km/s. The positions of these clusters on the plane of the sky, along with the rest of our globular cluster dataset, are illustrated in Figure \ref{fig:radec}. Figure \ref{fig:radec} illustrates that the majority of globular clusters, including 5 of the outermost 6 clusters, are found in the Southern sky ($\rm Dec \leq 0$). This lack of symmetry further exemplifies that the Galactic globular cluster population is incomplete, as the population should be spherically symmetric. The orbital properties of the outermost 6 clusters are listed in Table \ref{table:outergcs}, where $r_p$ and $r_{apo}$ are determined using the \citet{Irrgang13} potential. None of the 6 outermost clusters are believed to share an accretion origin \citep{Massari19}.

\begin{table*}
\centering
\begin{tabular}{lccccccccc}
Globular Cluster & RA & Dec & Distance & $\mu_{RA}$ & $\mu_{Dec}$ & $v_{los}$ & $r_p$ & $r_{apo}$ \\
{} & ($\rm degrees$) & ($\rm degrees$) &($\rm kpc$) & ($\rm mas \ yr^{-1}$) & ($\rm mas \ yr^{-1}$) & ($\rm km \ s^{-1}$) & (kpc) & (kpc) \\
\hline
Crater & 174.1 & -10.9 & 145.00 & 0.00 & -0.1 & 148.3 & 71.6 & 147.2 \\
AM1 & 58.8 & -49.6 & 123.30 & 0.02 & -0.2 & 118.0 & 77.6 & 143.7 \\
Pal4 & 172.3 & 29.0 & 103.00 & -0.21 & -0.4 & 72.4 & 11.8 & 108.5 \\
Pal3 & 151.4 & 0.1 & 92.50 & 0.08 & -0.1 & 94.0 & 59.0 & 113.7 \\
Eridanus & 66.2 & -21.2 & 90.10 & 0.40 & -0.1 & -23.8 & 30.5 & 136.4 \\
NGC2419 & 114.5 & 38.9 & 83.18 & -0.02 & -0.6 & -21.1 & 18.3 & 91.8 \\

\hline
\end{tabular}
\caption{The orbital properties of the 6 outermost globular clusters in the Milky Way}
\label{table:outergcs}
\end{table*}

\begin{figure}
    \includegraphics[width=0.48\textwidth]{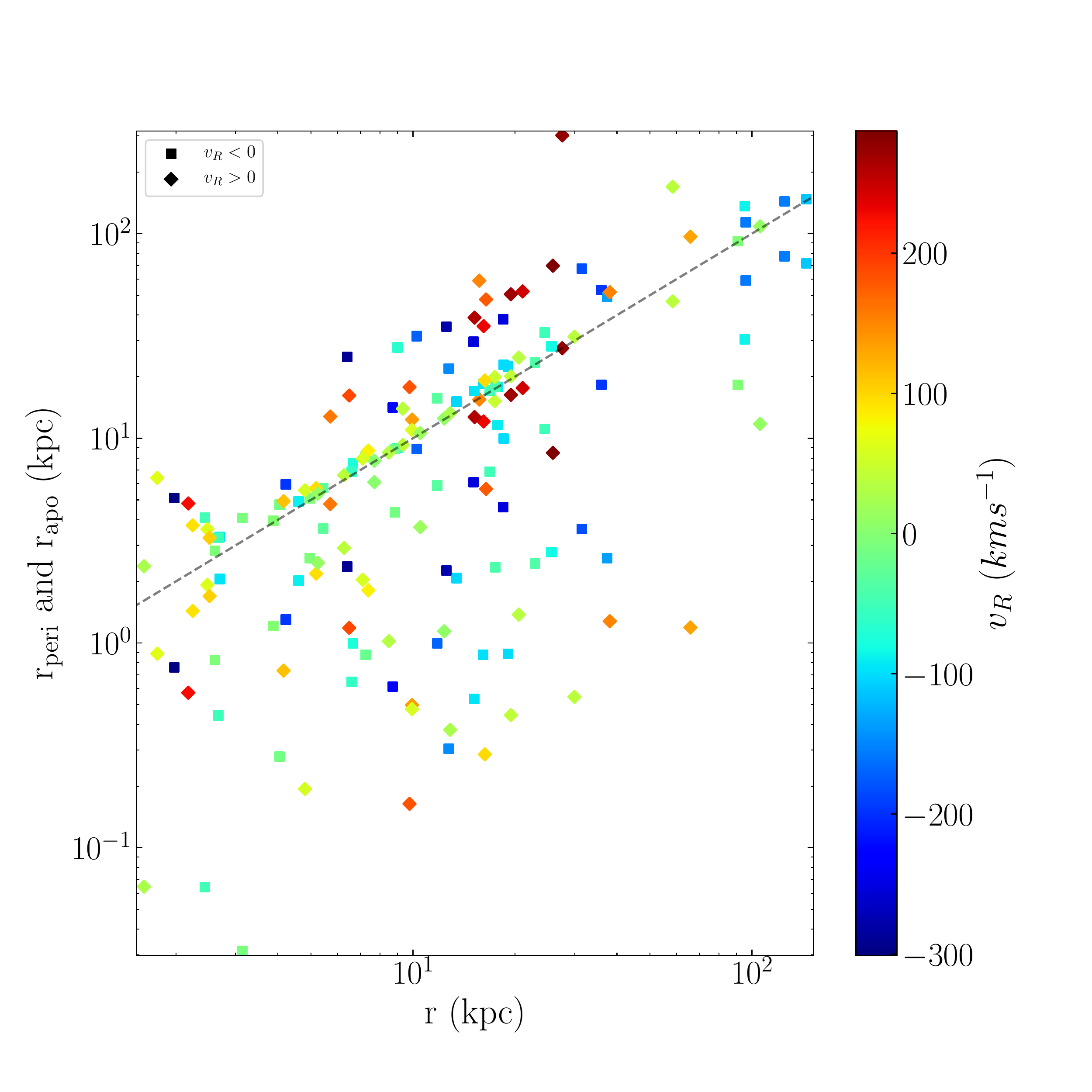}
    \caption{Current galactocentric distance $r$ of each Galactic cluster compared to both its pericentric and apocentric distances. Points are colour coded by the clusters current cylindrical radial velocity, with squares indicating a $v_R < 0$ and diamonds indicating $v_R > 0$. For comparison purposes, a 1:1 line is illustrated above which points correspond to apocentre and below which points correspond to pericentre.}
    \label{fig:vrplot}
\end{figure}

To look closer at the sharp decrease in the number of globular clusters beyond 50 kpc and a lack of low velocity clusters, Figure \ref{fig:vrplot} compares the current $r$ of each cluster to its pericentre and apocentre, with squares again corresponding to clusters moving inwards and diamonds corresponding to clusters moving outwards. Globular cluster orbits have been integrated with \texttt{galpy}\footnote{http://github.com/jobovy/galpy} \citep{Bovy15} using the \citet{Irrgang13} model for the Milky Way to remain consistent with \citet{Baumgardt19} catalogue. The colour bar marks each cluster's radial velocity. The figure clearly demonstrates our statement that most of the outer clusters are currently moving inwards, with the majority ($67\%$) of clusters beyond 30 kpc having $v_R < 0$. Figure \ref{fig:vrplot} also reveals that the several clusters beyond 30 kpc are currently quite far from their apocentres.

The uneven split of outer clusters with $v_R > 0$ and $v_R < 0$ serves as an indicator that there may be several undiscovered outer clusters with $v_R > 0$. Supporting this statement are the lack of outer clusters near their apocentres with near zero velocities and the asymmetric distribution of globular cluster declinations. The asymmetry also suggests a possible explanation for the cluster population being incomplete in that coverage of the sky is complete down to different limiting magnitudes. The next step is to consider how the cluster population evolves with time and if any further discrepancies in the global properties of the population exist.

\section{Orbit Averaging the Galactic Globular Cluster Population} \label{s_results}

\begin{figure}
    \includegraphics[width=0.48\textwidth]{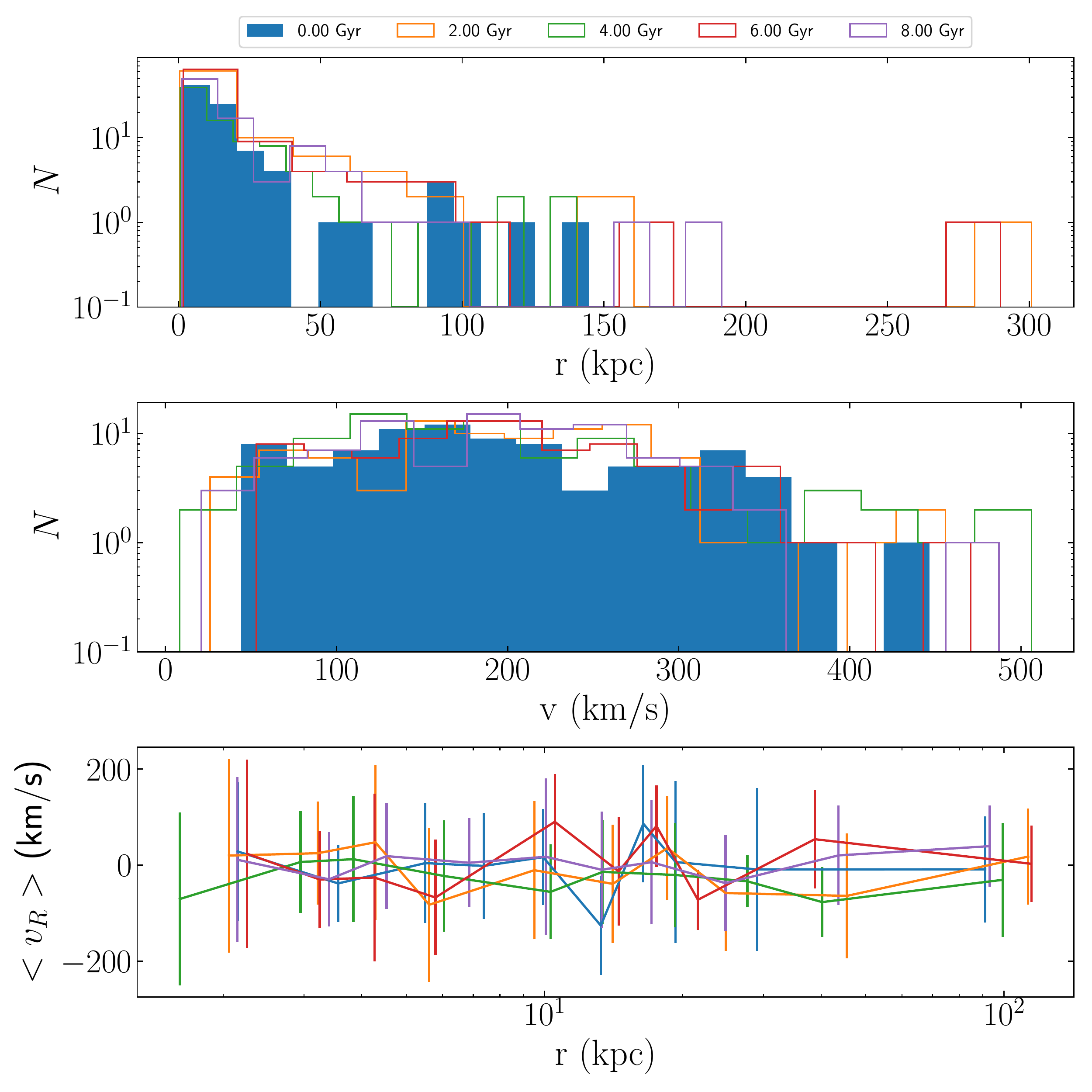}
    \caption{Static halo orbit integrations showing the variation of the radius distribution (top panel), total velocity distribution (middle panel), and cylindrical radial velocity profile (bottom panel) of the Galactic globular cluster population between present day and integrated for 12 Gyr.}
    \label{fig:hist_evol}
\end{figure}

We next consider how the present day spatial and kinematic distributions of Galactic clusters compare to how they may look at other times. We specifically study the distribution of they outer clusters as they move along their orbits in order to identify any discrepancies that may suggest the outer population is incomplete. The orbit of each globular cluster is integrated in a static potential for 12 Gyr using the \citet{Irrgang13} model for the Milky Way. The radial distribution, velocity distribution, and mean cylindrical radial velocity profile at times equal to 2, 4, 6, and 8 Gyr into the future are illustrated as representative times in Figure \ref{fig:hist_evol}. Note that integrating cluster orbits forwards to 12 Gyr is not meant to be an accurate representation of what the system looks like in 12 Gyr, as the evolution of the Milky Way itself and its satellites would need to be considered. The integration is performed in order to generate different configurations of the globular cluster system based on clusters being located at different phases in their orbits. 

The first thing to note in Figure \ref{fig:hist_evol} is that the radial distribution usually extends to larger radii than the currently known distribution, extending well beyond the distance to the outermost known cluster (Crater at 145 kpc). The radial distribution reaches its maximum extent at integration times 2 Gyr and 6 Gyr,  when NGC 4833 reaches an apocentre of 302 kpc. Furthermore, at integration times of 2 Gyr, 6 Gyr, and 8 Gyr there are several clusters between 145 kpc and 302 kpc. Only at the current epoch and at 4 Gyr is the Galactic globular cluster system constrained to within 145 kpc. Similarly, it is only at the current epoch that the system is sharply truncated at 50 kpc.

With respect to cluster kinematics, more clusters have lower velocities at the 4 Gyr, 6 Gyr, and 8 Gyr times, effectively `filling in' the low velocity end of the total velocity distribution as clusters travel towards their apocentres. It is interesting to note that, similar to the present day, at the 6 Gyr time epoch there are also no low-velocity clusters. Finally, the mean cylindrical radial velocity profiles also become more consistent with $\langle v_R \rangle =0$ km/s at 2 Gyr, 6 Gyr, and 8 Gyr than they are today. Hence the outer cluster population is not dominated by clusters moving inwards with $v_R<0$ as currently observed in the Milky Way and $\Delta {N_{v_R}}$ is closer to zero. 

\begin{figure*}
    \includegraphics[width=\textwidth]{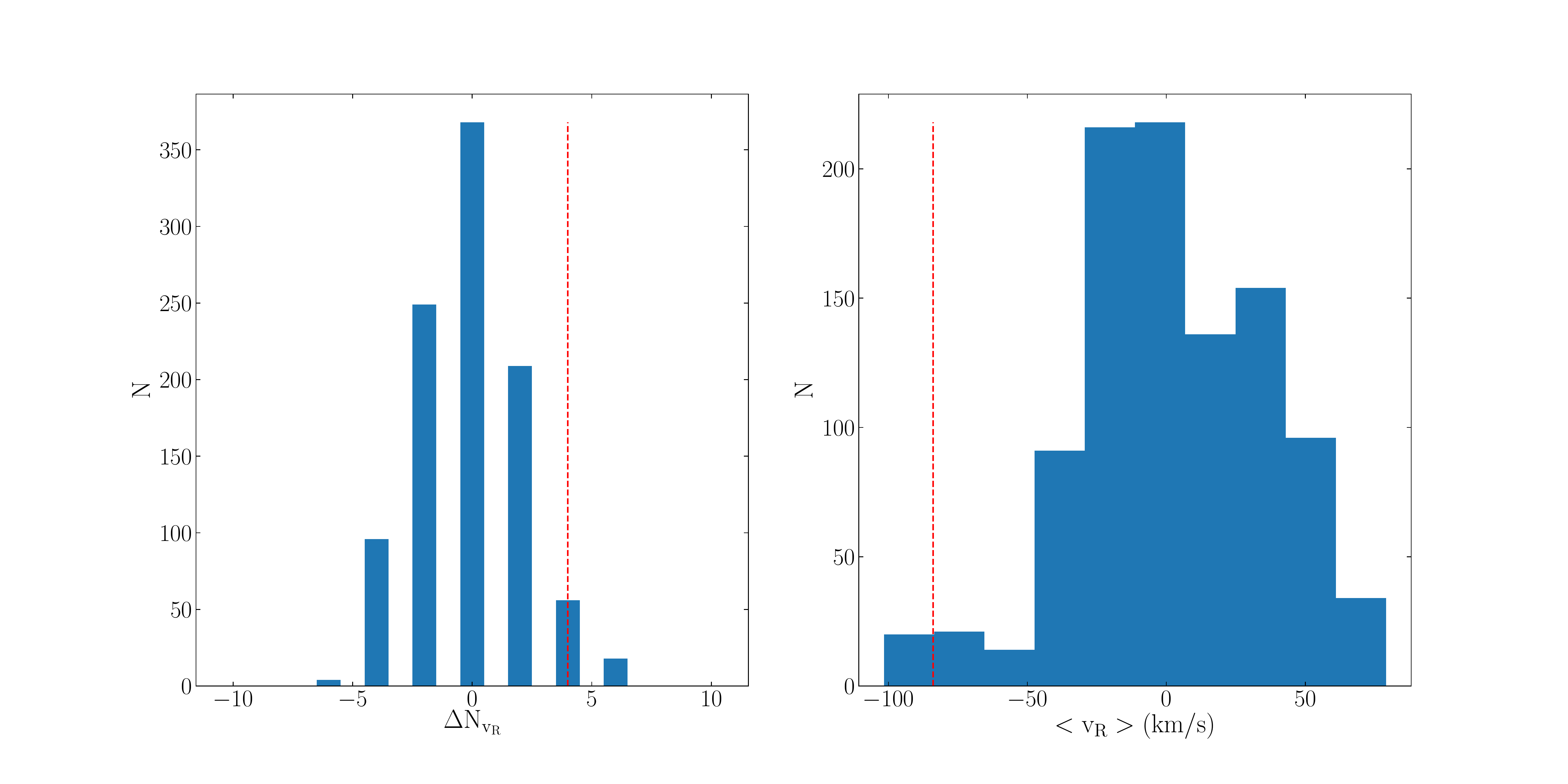}
    \caption{Left Panel: Difference between the number of the outer six clusters that are moving inwards and the number of the outer six clusters that are moving outwards $\Delta N_{v_R}$ in different configurations of the Milky Way globular cluster system. Right Panel: Mean cylindrical radial velocity $\langle v_R \rangle$ of the outer six globular clusters in different configurations of the Milky Way globular cluster system. The current properties of the outer six globular clusters in the Milky Way are shown as red dashed lines, indicating that Milky Way clusters are currently in a very rare configuration with a likelihood of $\sim 1.2\%$.}
    \label{fig:vrdis}
\end{figure*}

Taking a closer look at the properties of the outer population, Figure \ref{fig:vrdis} illustrates both $\Delta N_{v_R}$ and the mean cylindrical radial velocity $\langle v_R \rangle$ of the outermost six clusters in 1000 orbital configurations of the Milky Way's globular cluster system. For comparison purposes, the observed values are illustrated as dashed red lines. The distribution of difference in number of inward and outward moving clusters is modelled a binomial distribution with an equal chance of moving in or out. In only $6.1 \%$ of the configurations does the Milky Way have $\Delta N_{v_R}$ equaling 4, suggesting the Milky Way's current configuration is a low probability realization if the known outer clusters are a complete, equilibrium, distribution. Furthermore only $2\%$ of configurations have a $\langle v_R \rangle$ between -91 km/s and -72 km/s, which would be consistent the current $\langle v_R \rangle$ of -83.4 km/s. Combining these two factors, the percentage of configurations consistent with the Milky Way being in its current state is only $1.2\%$. 

We find that the current Galactic globular cluster population is less extended, lacking low-velocity clusters, and has more outer clusters moving inwards than the orbit averaged properties of the population. The discovery of more distant globular clusters that are currently moving outwards towards their apocentre would fill in the outer part of the radial profile and the low-velocity end of the total velocity distribution, while making the mean radial velocity of outer clusters and $\Delta {N_{v_R}}$ consistent with zero. Therefore it is necessary to next estimate the likelihood that there are no undiscovered and the Galactic cluster system is just currently in a very rare configuration and the likelihood that there exists one or more undiscovered globular clusters that would account for the observed discrepancies. An alternative explanation is of course that the Milky Way globular cluster system is currently not in equilibrium. However, given the most recent major globular cluster accretion event was the Sagittarius Dwarf Galaxy approximately 7 Gyr ago \citep{Ibata1994,Kruijssen20} (bringing with it approximately 6 clusters \citep{Massari19}), most clusters have undergone several orbits. The orbital period at 100 kpc is about 2 Gyr in the Milky Way and the fairly widely dispersed clusters should have spread out in velocity.

\section{The Likelihood of Undiscovered Galactic Globular Clusters} \label{s_likelihood}

For low-latitude clusters, the number of clusters between the Sun and the Galactic centre are compared to the number of clusters beyond the Galactic centre to estimate the likelihood of there being undiscovered low-latitude clusters. In a comparable approach, we use the difference $\Delta N_{v_R}$ between the number of outer clusters moving inwards ($N_{v_R}<0$) and the number of clusters moving outwards $N_{v_R}>0$ to estimate the likelihood of there being undiscovered outer clusters. More specifically, in order to further explore the completeness of the outer Galactic globular cluster system, we estimate the likelihood of there being $N_{extra}$ clusters given the current measurement of $\Delta N_{v_R}=4$. This type of estimate is ideally suited for a Bayesian analysis of the form:  

\begin{equation}\label{eqn:bayes}
P(A|B) = \frac{P(B|A) P(A)}{P(B)}
\end{equation}

where A is the event that $N_{extra}$ clusters exist in the outskirts of the Milky Way and B is the event that $\Delta N_{v_R}=4$ is found in the observed dataset.
Hence it is first necessary for us to determine the likelihood of measuring $\Delta N_{v_R}=4$ for a given $N_{extra}$ (P(B|A)), the probability of their being $N_{extra}$ undiscovered clusters (P(A)), and the probability of measuring $\Delta N_{v_R}=4$ (P(B)). Hence for our purposes, Equation \ref{eqn:bayes} can be written as:

\begin{equation}\label{eqn:prob}
P(N_{extra}|\Delta N_{v_R}=4) = \frac{P(\Delta N_{v_R}=4|N_{extra}) P(N_{extra})}{P(\Delta N_{v_R}=4)}
\end{equation}

\noindent It is important to note that estimating $P(N_{extra}|\Delta N_{v_R}=4)$ is independent of our choice in coordinate system.  Performing the analysis in spherical coordinates, $P(N_{extra}|\Delta N_{v_r}=4)$ will be the same as $P(N_{extra}|\Delta N_{v_R}=4)$ since the number of clusters with $v_R < 0$ will always be the same as the number of clusters with $v_r < 0$.

\begin{figure}
    \includegraphics[width=0.48\textwidth]{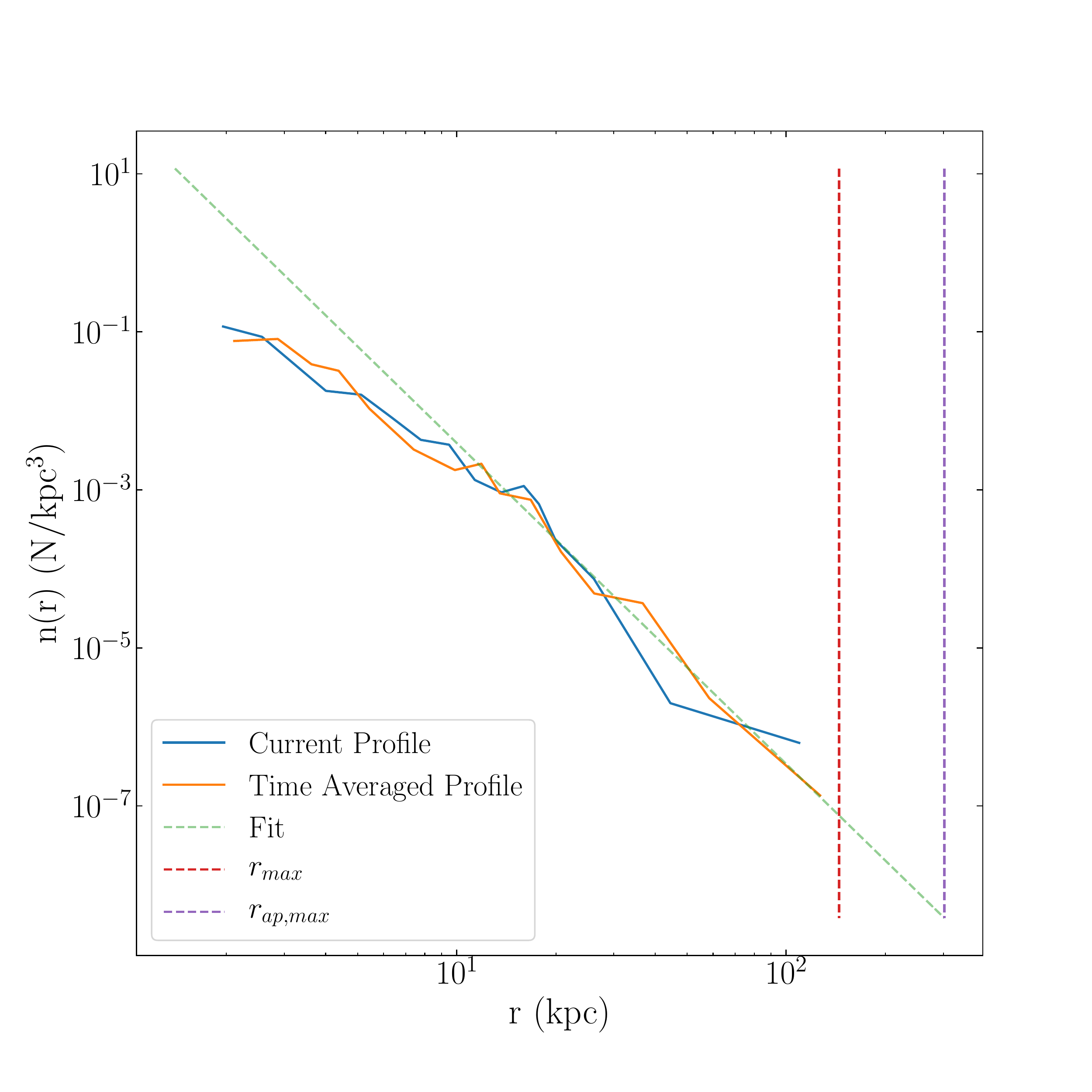}
    \caption{Current and time averaged number density profile of the Galactic globular cluster population. For illustrative purposes we also show a linear fit to the time average profile beyond 10 kpc. The vertical lines mark the current maximum radius and maximum apocentre of the population.}
    \label{fig:mwgc_evol}
\end{figure}

First, to estimate the probability of their being $N_{extra}$ undiscovered clusters ($P(N_{extra})$), we determine the time averaged number density profile of the Galactic globular cluster system ($<n(r)>$) using the 1000 randomly selected configurations that were determined from the orbit averaging performed in Section \ref{s_results}. The time-averaged profile, illustrated in Figure \ref{fig:mwgc_evol} alongside the current number density profile, is well represented by a power-law beyond 10 kpc. The best-fit power-law model, illustrated as a dashed green line, is of the form $\langle n(r) \rangle \propto r ^{-3.8 \pm 0.1}$. The fit is only slightly shallower than the current number density profile (see Figure \ref{fig:mwgc}), but has a lower uncertainty. For comparison purposes, the current $r_{max}$ and $r_{apo,max}$ are illustrated as vertical red and violet lines respectively.

As discussed above, the observed Galactic globular cluster number density profile drops sharply beyond 50 kpc. Therefore it is a reasonable first assumption that the high-latitude Galactic globular cluster population the completeness radius, ($r_{complete}$), may be as small as 50 kpc. Integrating the time averaged number density profile between  between 50 kpc and $r_{apo,max}$ results in an estimate of there being $7.1 \pm 5.4$ clusters beyond 50 kpc. The uncertainty is estimated by first performing the integration over 1000 test cases, where $<n(r)>$ is varied within the uncertainty in the power-law fit and then taking the dispersion of the distribution. The result is consistent with the number of observed high-latitude clusters beyond 50 kpc, which is 8. 

We also consider the more conservative cases of $r_{complete} = 100$ kpc and $ 150$ kpc, where integrating the time averaged number density profiles finds there should be $2.8 \pm 2.4$ clusters and $1.4 \pm 1.2$ clusters, respectively. The number of observed clusters beyond 100 kpc and 150 kpc are 3 and 0 respectively. For a given $r_{complete}$, the integral is always consistent with the observed number of clusters, as expected, given that the observed clusters are used to measure the number density profile to begin with. It is, however, the uncertainty in the fit that supports the possibility of there being 1 or more additional clusters yet to be discovered.

\begin{figure*}
    \includegraphics[width=\textwidth]{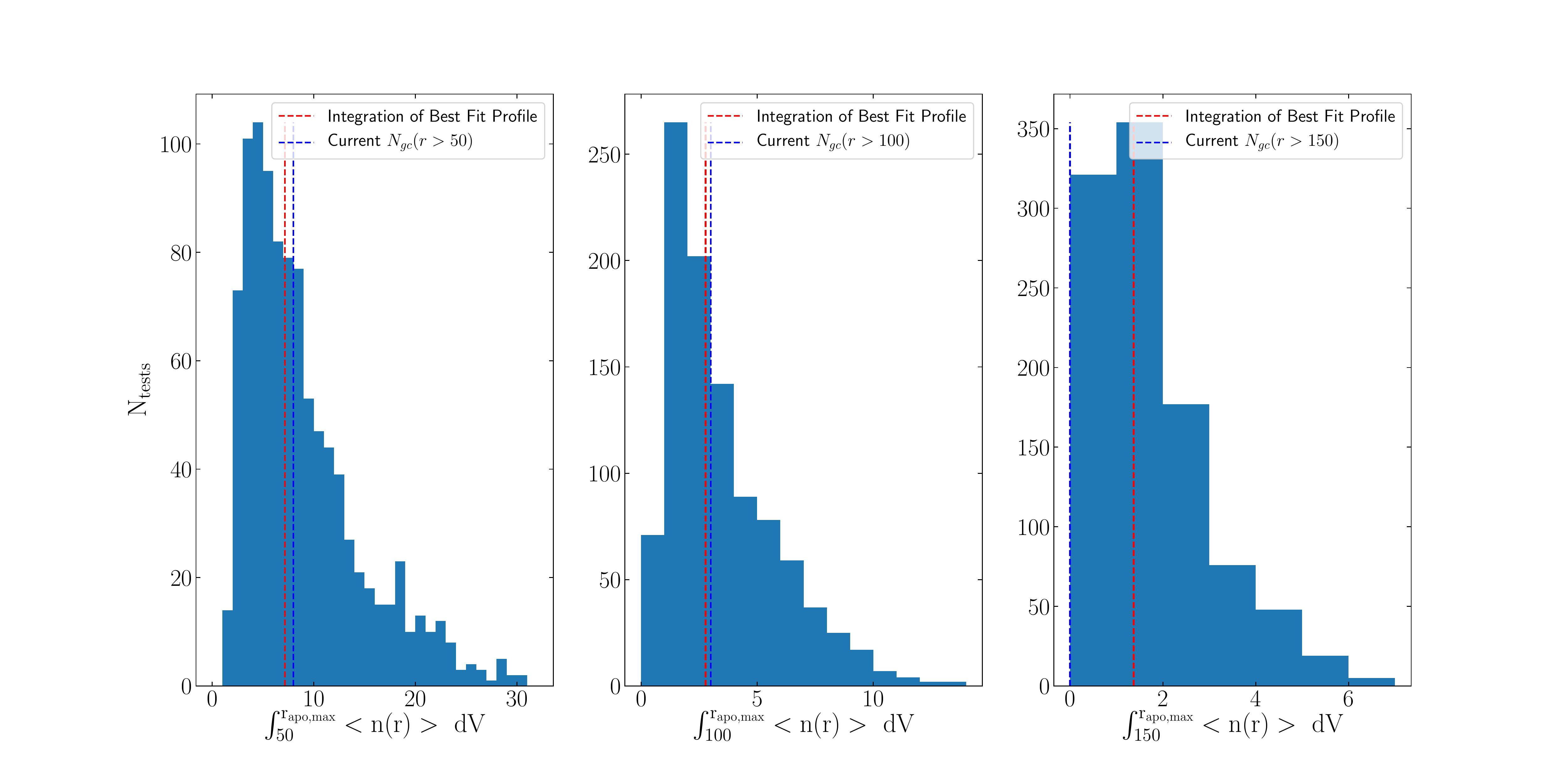}
    \caption{Integral of the power-law fit to the time-averaged number density profile of the Galactic globular cluster population, where the fit has been randomly sampled within uncertainty for $r_{complete}$ = 50 kpc (left panel), 100 kpc (middle panel), and 150 kpc (right panel). P($N_{extra}$) is taken to be the fraction of cases where the difference between the integral and the observed number of clusters is $N_{extra}$. P(0) is taken to be the fraction of cases where the difference is less than or equal to zero.}
    \label{fig:rho_ints}
\end{figure*}

The full distribution of estimates for the number of clusters beyond 50, 100, and 150 kpc are illustrated in Figure \ref{fig:rho_ints}, along with the estimate from the best-fit to the time-averaged profile and the currently observed value. Figure \ref{fig:rho_ints} can then be used to estimate $P(N_{extra})$ in each case by finding the fraction of cases where the difference between the integral and the observed number of clusters beyond $r_{complete}$ is $N_{extra}$. For $P(N_{extra})=0$, we sum all test cases that return a value less than or equal to the current number of clusters beyond $r_{complete}$. The probabilities for $N_{extra}$ between 0 and 6 are listed in Table \ref{table:probs}, where we also note the fraction of cases with number density profiles that suggest more than 6 extra clusters.

\begin{table}
\centering
\begin{tabular}{l|ccc}
$N_{extra}$ & {} & $P(N_{extra})$ & {} \\
\hline
 $r_{complete}$ &   50 kpc &  100 kpc &  150 kpc \\
\hline
0 & 0.61 & 0.72 & 0.36 \\
1 & 0.06 & 0.09 & 0.32 \\
2 & 0.04 & 0.07 & 0.17 \\
3 & 0.04 & 0.04 & 0.08 \\
4 & 0.04 & 0.02 & 0.04 \\
5 & 0.03 & 0.03 & 0.02 \\
6 & 0.02 & 0.01 & 0.01 \\
>6 & 0.15 & 0.03 & 0.00 \\

\hline
\end{tabular}
\caption{Probability of $N_{extra}$ undiscovered Galactic globular clusters for completeness radii of 50 kpc, 100 kpc, and 150 kpc. These values are used for $P(N_{extra})$ in Equation \ref{eqn:prob}}
\label{table:probs}
\end{table}

We note that an alternative approach would be to only fit $\langle n(r) \rangle$ between 10 kpc and $r_{complete}$, as this method avoids using incomplete regions when measuring $<n(r)>$. However, due to the lower number of clusters and the narrower range in $r$ when $r_{complete} =$ 50 kpc and 100 kpc, the uncertainty in the fits to $\langle n(r) \rangle$ will be much larger. Integrating $\langle n(r) \rangle$ between $r_{complete}$ and $r_{apo,max}$ will then result in predicting that even more clusters exist beyond $r_{complete}$ than when all clusters beyond 10 kpc are used to measure $\langle n(r) \rangle$. Our approach is therefore more conservative, such that our estimates of $P(N_{extra}>0)$ in Table \ref{table:probs} can be treated as lower limits.

For all three values of $r_{complete}$, the most probable scenario is that there are no additional globular clusters. However this estimate does not take into consideration the anomalous fact that the outer six globular clusters have $\Delta N_{v_R}=4$. The likelihood of measuring $\Delta N_{v_R}=4$ for a given $N_{extra}$ ($P(\Delta N_{v_R}=4|N_{extra})$) can be estimated from orbit averaging experiments that are similar to the one performed in Section \ref{s_results}. In fact Figure \ref{fig:vrdis} already provides $P(\Delta N_{v_R}=4|N_{extra}=0)=6.1\%$. For $N_{extra}>0$, artificial globular clusters need to be generated.

To orbit average Galactic globular cluster populations with $N_{extra}$ extra clusters, we first generate 1000 artificial clusters with positions based on the time averaged number density profile and cylindrical coordinate velocities from Gaussian distributions with dispersions equal to our observed dataset. We assume the undiscovered clusters have a current radius between $r_{complete}$ and $r_{apo,max}$. When orbit averaging, where we find the $\Delta N_{v_R}$ distribution of the population at random timesteps, $N_{extra}$ clusters are randomly selected from the population of artifical clusters. It is also important to note that in each case we actually measure $\Delta N_{v_R}$ for the outer $6 + N_{extra}$ clusters. The orbit averaging technique is performed for $0 \le N_{extra} \le 6$ and $r_{complete}=50, 100, 150$ kpc.

\begin{figure*}
    \includegraphics[width=\textwidth]{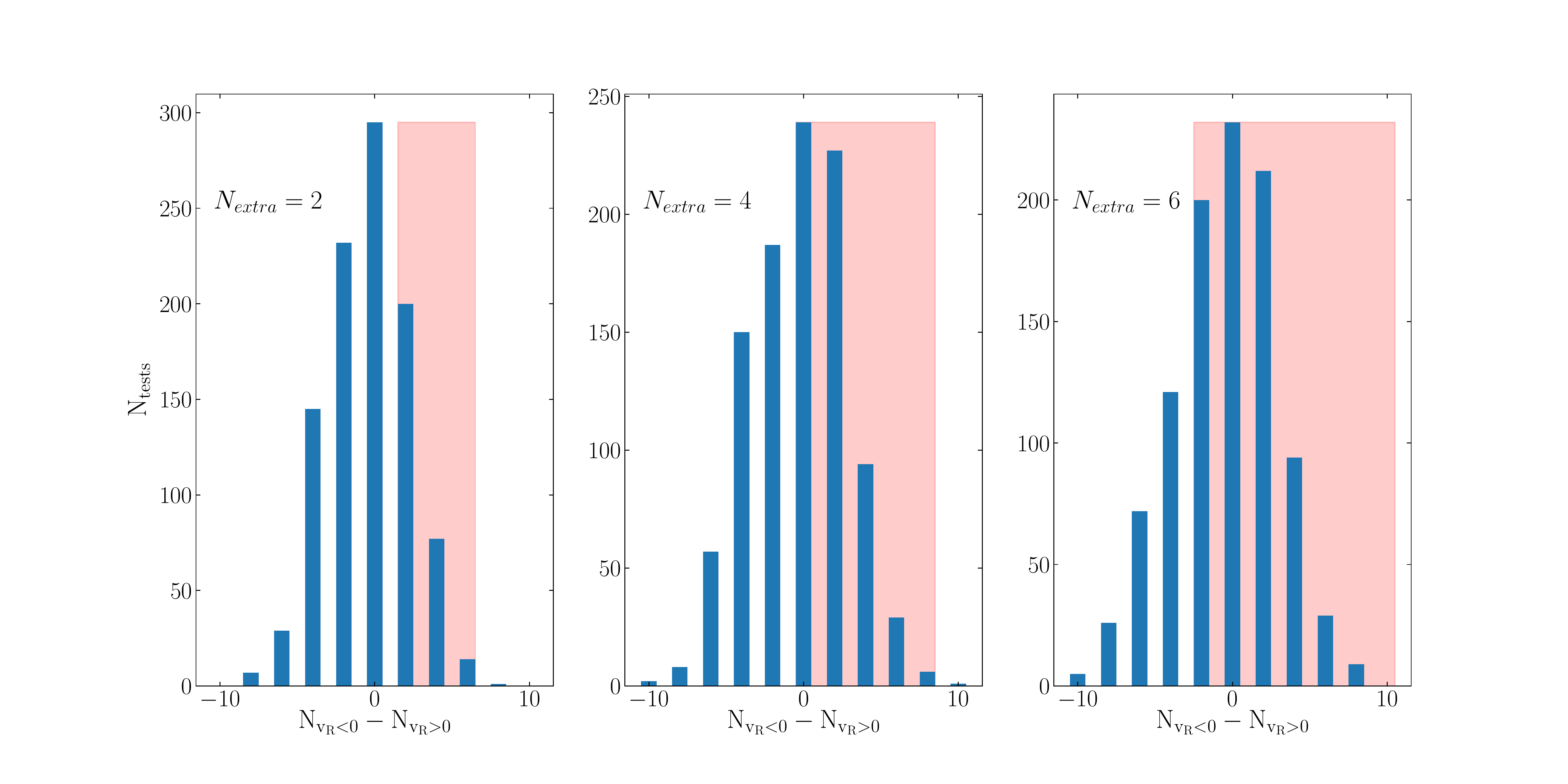}
    \caption{Difference between the number of the outer six + $N_{extra}$ clusters that are moving inwards and the number of the outer six + $N_{extra}$ clusters that are moving outwards $\Delta N_{v_R}$ in different configurations of the Milky Way globular cluster system. $N_{extra}$ artificially generated clusters are inserted into the Galactic globular cluster population for $N_{extra}=2$ (left panel), $N_{extra}=4$ (middle panel), and $N_{extra}=6$ (right panel and $r_{complete}=50$ kpc). The allowable range of $\Delta N_{v_R}$ values in the Milky Way, given that $\Delta N_{v_R}$ of the six outermost known clusters is 4, are in the shaded red region. The fraction of cases within the allowable range is used to determine $P(\Delta N_{v_R}=4|N_{extra}$ in Equation \ref{eqn:prob}.}
    \label{fig:dv_ints}
\end{figure*}

As an example, Figure \ref{fig:dv_ints} illustrates the $\Delta N_{v_R}$ distribution for cases where 2, 4, and 6 artificial clusters have been added to the Galactic globular cluster population assuming $r_{complete}=50$ kpc. For each value of $N_{extra}$, the range of $\Delta N_{v_R}$ values that could be reached by the Milky Way are highlighted by the red shaded region. The fraction of cases in the red shaded region therefore represent $P(\Delta N_{v_R}=4|N_{extra})$, as one would still measure $\Delta N_{v_R}=4$ if $N_{extra}$ clusters have yet to be observed. For example the right panel illustrates that if $N_{extra}=6$, it is possible to measure $\Delta N_{v_R}=4$ for 6 outer clusters while the true $\Delta N_{v_R}$ of the outermost 6+$N_{extra}$ clusters could be anywhere between -1 and 10, with 0 extra being the most probable case.

\begin{figure}
    \includegraphics[width=0.48\textwidth]{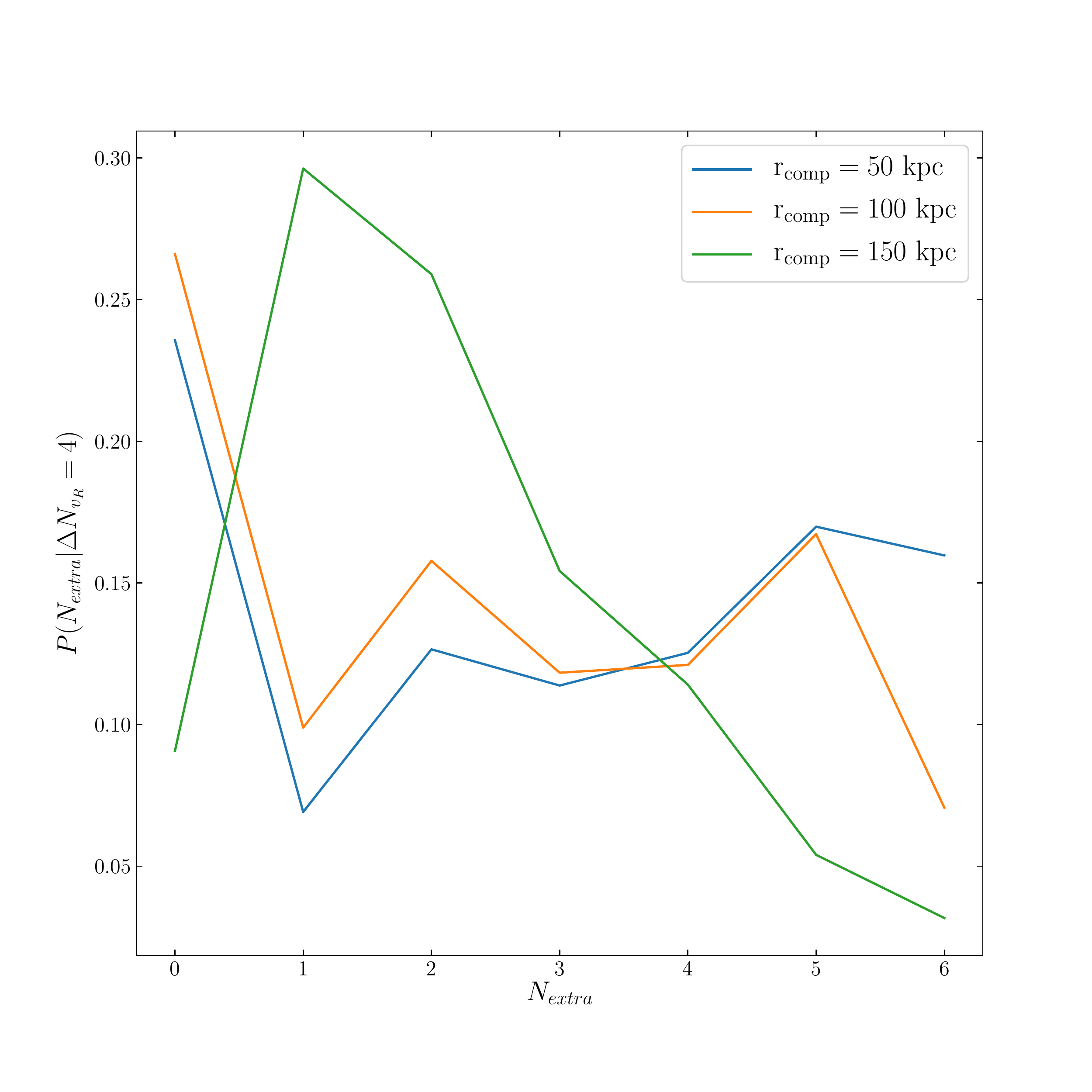}
    \caption{Likelihood of there being $N$ undiscovered globular clusters in the Milky Way given that we observe $\Delta N_{v_R}=4$ ($P(N_{extra}|\Delta N_{v_R}=4)$ in Equation \ref{eqn:prob}) for $r_{complete} = 50, 100, 150$ kpc.}
    \label{fig:likeplot}
\end{figure}

Taking $P(\Delta N_{v_R}=4)$ in Equation \ref{eqn:prob} to be the sum of $P(\Delta N_{v_R}=4|N_{extra}) P(N_{extra})$, we are now able to calculate the likelihood of there being $N_{extra}$ undiscovered clusters in the outskirts of the Milky Way ($P(N_{extra}|\Delta N_{v_R}=4)$) using Equation \ref{eqn:prob}. The results of this calculation for all three cases of $r_{complete}$ are illustrated in Figure \ref{fig:likeplot} and listed in Table \ref{table:likes}.

\begin{table}
\centering
\begin{tabular}{l|ccc}
$N_{extra}$ & {} & $P(N_{extra}|\Delta N_{v_R}=4)$ & {} \\
\hline
 $r_{complete}$ &   50 kpc &  100 kpc &  150 kpc \\
\hline
0 & 0.24 & 0.27 & 0.09 \\
1 & 0.07 & 0.10 & 0.30 \\
2 & 0.13 & 0.16 & 0.26 \\
3 & 0.11 & 0.12 & 0.15 \\
4 & 0.13 & 0.12 & 0.11 \\
5 & 0.17 & 0.17 & 0.05 \\
6 & 0.16 & 0.07 & 0.03 \\

\hline
\end{tabular}
\caption{Likelihood of $N_{extra}$ undiscovered Galactic globular clusters in the Milky Way given the fact that $\Delta N_{v_R}=4$ is found in the observed dataset ($P(N_{extra}|\Delta N_{v_R}=4)$ in Equation \ref{eqn:prob}). We consider cases where the $N_{extra}$ clusters are currently found beyond a completeness radii of 50 kpc, 100 kpc, and 150 kpc.}
\label{table:likes}
\end{table}

Taking into consideration both the time-averaged number density profile of the high-latitude Galactic globular cluster system and the kinematics of the outer clusters, we find that the likelihood of the system being complete (i.e. no undiscovered clusters) is between $10\%$ and $27\%$, depending on whether the system is currently complete out to 50 kpc, 100 kpc, or 150 kpc. Hence if the observed cluster distribution is complete out to either 50 kpc or 100 kpc, there is a $75\%$ of there being at least one undiscovered globular cluster in the outer regions of the Milky Way. If the system is complete out to 150 kpc (just beyond the orbit of the outermost cluster Crater), the likelihood jumps to $90\%$ that there is at least one undiscovered cluster. The likelihood of there being two undiscovered clusters drops below $70\%$ for all three $r_{complete}$ considered. The exact likelihoods of there being at least one undiscovered cluster ($P(N_{extra} >= 1|\Delta N_{v_R}=4)$) and at least two undiscovered clusters ($P(N_{extra} >= 2|\Delta N_{v_R}=4)$) are listed in Table \ref{table:likes_compare}. Also listed in Table \ref{table:likes_compare} are the likelihoods calculated with various exlcusions and additions to our dataset, which we discuss in the following sections.

\begin{table*}
\centering
\begin{tabular}{l|cccccc}
Restrictions & {} & $P(N_{extra} >= 1|\Delta N_{v_R}=4)$ & {} & {} & $P(N_{extra} >= 2|\Delta N_{v_R}=4)$ & {} \\
\hline
 $r_{complete}$ &   50 kpc &  100 kpc &  150 kpc &   50 kpc &  100 kpc &  150 kpc \\
\hline
None & 0.76 & 0.73 & 0.91 & 0.69 & 0.63 & 0.61 \\
Exclude $b\le20$ & 0.8 & 0.85 & 0.97 & 0.73 & 0.73 & 0.78 \\
Include Koposov 1 and 2 & 0.78 & 0.81 & 0.94 & 0.70 & 0.70 & 0.70 \\
\citet{Bovy15} Potential & 0.79 & 0.82 & 0.91 & 0.72 & 0.7 & 0.66 \\
\citet{mcmillan07} Potential & 0.61 & 0.68 & 0.81 & 0.55 & 0.60 & 0.62 \\
\hline
\end{tabular}
\caption{Likelihood of $N_{extra} \ge 1$ and $N_{extra} \ge 2$ undiscovered Galactic globular clusters in the Milky Way given the fact that $\Delta N_{v_R}=4$ is found in the observed dataset ($P(N_{extra}|\Delta N_{v_R}=4)$ in Equation \ref{eqn:prob}). We consider cases where the $N_{extra}$ clusters are currently found beyond a completeness radii of 50 kpc, 100 kpc, and 150 kpc. The first row lists likelihoods calculated using our main dataset, with subsequent rows representing likelihoods calculated with modified datasets or Galactic potentials. The main dataset represents the most conservative estimates of $P(N_{extra} \ge 1|\Delta N_{v_R}=4)$ and $P(N_{extra} \ge 2|\Delta N_{v_R}=4)$.}
\label{table:likes_compare}
\end{table*}

\subsection{Low Latitude Completeness}\label{s_latcomplete}

One concern that was first discussed to in Section \ref{s_method} is that the globular cluster population is incomplete at low Galactic latitudes due to reddening and extinction. In order to ensure a complete Galactic globular cluster dataset, we did not include clusters with $b<10^{\circ}$. It is worthwhile, however, to test whether this choice in $b$ is conservative enough. We therefore repeat the above analysis but instead using globular clusters with $b>20^{\circ}$, which leaves us with only 49 clusters to analyze. It is important to note that NGC 5139 (Omega Cen) is also removed from the dataset when testing low latitude completeness, effectively exploring whether or not its inclusion in the dataset affects the results.

With this high latitude  dataset the Galactic globular clusters number density profile is shallower, which increases the probability of their being undiscovered clusters at large galactocentric radii. For example, our estimates for $P(0)$ decrease to $\sim 40\%$ for $r_{complete} = $ 50 and 100 kpc. For $r_{complete} = $ 150, $P(0)$ becomes zero. Even $P(1)$ is lower for all three values of  $r_{complete}$ than the values presented in Table \ref{table:probs}. The additional exclusion of clusters with $10^{\circ}< b <20^{\circ}$ does little to $P(\Delta N_{v_R}=4|N_{extra})$ and $P(\Delta N_{v_R}=4)$ since these low-latitude clusters all have relatively small apocentres. Hence they never affect the calculation of $\Delta N_{v_R}$.

For all three values of $r_{complete}$, the likelihood of there being at least one undiscovered cluster increases to at least $94\%$. For  $r_{complete} = $ 50 kpc the probability of there being two undiscovered clusters then drops to $80\%$ while it stays above $90\%$ for $r_{complete} = $ 100 and 150 kpc. The likelihoods of there being at least one and two undiscovered clusters for all three values of $r_{complete}$ using this restricted dataset are list in the second row of Table \ref{table:likes_compare}. In all cases these likelihoods are larger than reported in Table \ref{table:likes}, suggesting our initial latitude restriction results in a more conservative estimate of the number of undiscovered clusters.

It is worth noting that when using the entire list of 160 confirmed clusters in \citet{Vasiliev19} and \citet{Harris10}, the orbit averaged Galactic globular cluster number density profile is steeper than when only globular clusters with $b>10^{\circ}$ or $b>20^{\circ}$ are used. For $r_{complete} = 50$ kpc or 100 kpc, $P(0)$ is greater than $90\%$. For $r_{complete} = 150$ kpc, $P(0)$ is approximately $50\%$. While this population is almost certainly incomplete, it is worthwhile noting how the underlying cluster number density profile affects $P(N_{extra)}$.

\subsection{Koposov 1 and Koposov 2} \label{s_koposov}

Two clusters listed in \citet{Harris10} were omitted from out dataset because they do not have measured proper motions. It is reasonable to then question whether or not the inclusion of Koposov 1 and Koposov 2 in the analysis would affect our estimate that the likelihood of there being at least undiscovered clusters in the Milky Way is $\sim 75\%$ or higher. We address this possibility by generating 1000 different orbits for Koposov 1 and Koposov 2 and redoing our analysis, where we now select a different orbit for the two clusters at every time epoch. When generating random orbits for Koposov 1 and Koposov 2, cylindrical coordinate velocities are selected from Gaussian distributions centred around a mean velocity of zero with dispersions equal to the cylindrical velocity dispersions of the observed dataset.

When calculating $P(N_{extra})$ assuming $r_{complete} = 50$ kpc, including randomized orbits for Koposov 1 and Koposov 2 leads to $P(0)$ and $P(>6)$ having standard deviations on the order of $5\%$. All other estimates of $P(N_{extra})$ vary by $1.5\%$ or less. The influence of Koposov 1 and Koposov 2 is slightly more significant when considering $r_{complete} = 100$ kpc and $r_{complete} = 150$ kpc, with our estimates of $P(N_{extra})$ varying by up to approximately $6.5\%$ depending on which randomly selected orbit is used in the analysis. However, most estimates have standard deviations less than $3\%$. Hence including these two clusters in the measurement and integration of the Galactic globular cluster populations density profile has a minimal effect on our calculation of $P(N_{extra})$, as they do not significantly alter the best-fit number density profile.

In terms of $P(\Delta N_{v_R}=4|N_{extra})$ and $P(\Delta N_{v_R}=4$, in only $5\%$ of cases do the randomly generated orbits for either Koposov 1 or Koposov 2 result in them being one of the outer six Galactic globular clusters. Hence, despite their inclusion in the analysis, our calculated probabilities for $P(\Delta N_{v_R}=4|N_{extra})$ and $P(\Delta N_{v_R}=4)$ will remain largely unchanged. Therefore the likelihoods of there being $N_{extra}$ clusters presented in Table \ref{table:likes} are minimally affected by the inclusion of Koposov 1 and Koposov 2 in the analysis. The exact likelihoods of there being at least one and two undiscovered clusters for all three values of $r_{complete}$ when including Koposov 1 and Koposov 2 in the dataset are listed in the third row of Table \ref{table:likes_compare}.)

\subsection{Globular Cluster Candidates} \label{s_gccs}

While \citet{Harris10} and \citet{Baumgardt19} list all of the confirmed Galactic globular clusters, there are a number of recently proposed globular cluster candidates whose inclusion in our analysis may affect the results \citep{Ryu18}. In fact, several of these candidates could be the undiscovered globular clusters that we predict have a high likelihood of existing as they have large galactocentric distances. However at this stage, the inclusion of these candidates (or a subset) would be purely speculative without confirmation that they are indeed globular clusters. Furthermore, since these candidates have not had their three-dimensional velocities measured, their orbits would have to be randomly generated similar to when we included Koposov 1 and 2 in our analsysis in Section \ref{s_koposov}. Having nearly 1/4 of the dataset consisting of unconfirmed globular clusters with unknown velocities would introducing significant uncertainty into our prediction. We can, however, qualitative explore how the inclusion of these candidates would affect our results.

While \citep{Ryu18} lists $\sim$ 25 candidates that are not associated with the Galactic plane, we will focus specifically on the 8 ultra-faint objects that \citet{Contenta2017} find to be consistent with their prediction of their being $3^{+7.3}_{-1.6}$ faint star clusters beyond 20 kpc. Of these candidates, DES 1 \citep{Luque2016_des1}, Eridanus III \citep{Conn2018}, and Kim 2 \citep{Kim2015b} all have galactocentric distances greater than 85 kpc. If these candidates and confirmed as clusters and have positive radial velocities such that they are moving outwards, their inclusion in the dataset will result in $\Delta N_{v_R}$ decreasing towards zero. Such a change would place the Galactic globular cluster in a far more likely state than the currently observed $\Delta N_{v_R}=4)$. However if their radial velocities are such that $\Delta N_{v_R}$, the likelihood of their being $N_{extra}$ undiscovered cluster in the outer regions of the Milky Way will significantly increase.

Balbinot 1 \citep{Balbinot2013} and Mu{\~n}oz 1 \citep{Munoz2012} have galactocentric distances of $\sim$ 32 kpc and 45 kpc respectively, making them comparable to Koposov 1 and Koposov 2. Hence their inclusion in our analysis, without known orbits, may slightly affect $P(N_{extra}$ as $<n(r)>$ will decrease at a shallower rate and increase the predicted number of outer clusters. Overall, however, their inclusion will have a minimal affect on $P(N_{extra}|\Delta N_{v_R}=4)$ as they are most likely not outer clusters that are simply near pericentre.

Segue 3 \citep{Belokurov2010}, Kim 1 \citep{Kim2015}, and Kim 3 \citep{Kim2016} on the other hand have galactocentric distances less than 20 kpc. Hence their inclusion would result in the central time-averaged number density profile to increase such that $<n(r)>$ will decrease at a stepper rate and decrease the predicted number of outer clusters. Such a change may decrease $P(N_{extra}|\Delta N_{v_R}=4)$ as well if only these three candidates are added to the dataset. If Balbinot 1 and Mu{\~n}oz 1 are included as well, the net change to $<n(r)>$ may be negligible.

It is also worth noting that locations of the 8 ultra-faint objects are evenly split amongst the Northern and Southern skies. Hence including all 8 candidates as globular clusters does not remedy the asymmetric distribution of cluster declinations. In fact the three outermost candidates all have $\rm Dec < 0$. Therefore it still remains likely that the Galactic globular cluster is incomplete and limiting magnitude down to which different regions of the sky is complete needs to be considered when search for undiscovered globular clusters.

\subsection{Dependence on Galactic Potential}

Given that our results are based on the orbit averaging of Galactic globular clusters, it is likely that our conclusions depend strongly on our choice for the potential of the Milky Way. While our initial choice was the \citet{Irrgang13} to remain consistent with the course of our data \citet{Baumgardt19}, it is worthwhile to consider other commonly used potentials. Therefore we again repeat our analysis using the \texttt{MWPotential2014} galaxy model from \citet{Bovy15} and the \citet{McMillan17} model, both available in \texttt{galpy}. Note that clusters AM 4 and Laevens 3 are not bound to \texttt{MWPotential2014} and AM 4 is not bound to the \citet{McMillan17} model, given the most recent estimates of their proper motions. Hence they are not included in the analysis.  

The orbit-averaged Galactic globular cluster number density profile is shallower when orbits are integrated in \texttt{MWPotential2014} compared to our assumed model from \citet{Irrgang13}.  Beyond 10 kpc, the profile is well represented by a power-law of the form $<n(r)> \propto r ^{-3.5 \pm 0.2}$. Hence we expect $P(N_{extra})$ to be larger when using the \texttt{MWPotential2014} model, as integrating $<n(r)>$ from $r_{complete}$ to $r_{apo,max}$ will return a larger estimate for the number of stars beyond $r_{complete}$. For example, assuming $r_{complete}$ = 50 kpc, using \texttt{MWPotential2014} suggests that $12.5 \pm 10.0$ clusters exist beyond 50 kpc while we estimated $7.1 \pm 5.4$ using the \citet{Irrgang13} model. Comparing to Table \ref{table:probs}, $P(0)$ decreases by a factor of $1/2$ for $r_{complete} =$ 50 and 100 kpc and a factor of $1/3$ for $r_{complete} =$  150 kpc. In all cases, $P(0<(N_{extra}<6)$ increases by a few percent, while $P((N_{extra}>6)$ increases by on the order of $20\%$ for $r_{complete} =$ 50 and 100 kpc. The corresponding change to $P(N_{extra}|\Delta N_{v_R}=4)$ is that using \texttt{MWPotential2014} increases the likelihood of there being at least one undiscovered globular cluster in the Milky Way by $3\%$ and $9\%$ for $r_{complete} =$ 50 and 100 kpc. No change is seen for $r_{complete} =$ 150, which is not surprising given that only a small number of clusters have orbits that bring them beyond 150 kpc. See the fourth row of Table \ref{table:likes_compare} for the exact values of $P(N_{extra}|\Delta N_{v_R}=4)$ for all three value of $r_{complate}$ when using \texttt{MWPotential2014} as the tidal field of the Milky Way.

If we instead assume the \citet{McMillan17} model for the Milky Way, $\langle n(r) \rangle$ is nearly identical to when we use the \citet{Irrgang13} model with $<n(r)> \propto r ^{-3.8 \pm 0.3}$. The only difference between the two fits is that that the uncertainty in the fit to  $\langle n(r) \rangle $ is slightly larger when assuming the \citet{McMillan17} model. Hence differences in the calculation of $P(N_{extra})$ are minimal, as shown in the last row of Table \ref{table:likes_compare}.

\section{Discussion and Conclusions} \label{s_conclusion}

An analysis of the current and orbit averaged Galactic globular cluster population finds a likelihood of at least $73\%$ that there exists at least one undiscovered cluster in the outer regions of the high galactic latitude Milky Way. This estimate assumes that the system is in equilibrium and that the time-averaged outer number density profile  $\langle n(r) \rangle $ is adequately represented with a power-law. The likelihood could be as high as $91\%$, depending on the galactocentric distance out to which the observed population is currently complete. The estimate is based on the fact that of the outer six Galactic globular clusters, five have $v_R<0$ and only one has $v_R > 0$ ($\Delta N_{v_R}$=4), which is a rare configuration for a system in equilibrium. The likelihood of there being at least two undiscovered globular clusters is between $60\%$ and $70\%$, again depending on the completeness of the current population. The likelihood of there being three undiscovered clusters is on the order of $50\%$ or lower. The discovery of 1-3 additional clusters that are moving outwards would further align the present day cylindrical radial velocity distribution and total velocity distribution of Galactic clusters with their time-averaged distributions. 

These estimates rely on  a subset of the Galactic globular cluster population that is complete within the a given galactocentric distance $r_{complete}$. For example, incompleteness at low Galactic latitudes could lead to incorrect estimates of $\langle n(r) \rangle $  or $\Delta N_{v_R}$ if any of the clusters have large apocentres. We attempt to circumvent the effects of completeness by primarily focusing on clusters with $b > 10^{\circ}$, but study how completeness affects our prediction by performing our analysis using only clusters with $b > 20^{\circ}$. We also consider other models of the potential of the Milky Way and the inclusion of globular clusters with unknown kinematic properties, as both affect $\langle n(r) \rangle $. In all cases, our assumed latitude cutoff ($10^{\circ}$) and galaxy model \citep{Irrgang13} produce the most conservative estimates for the likelihood of there being $N_{extra}$ undiscovered globular clusters in the Milky Way. It is important to note, however, that even with these conservative restirctions there is a larger number of clusters with $\rm Dec < 0$ than $\rm Dec > 0$. Hence its possible that assuming the population is complete within 50 kpc is incorrect, and that many more clusters exist with $\rm Dec > 0$ than we predict here.

The mean mass, half-light radius, and absolute V magnitude of known Galactic clusters with $r>30$ kpc are $1.7 \times 10^5 M_{\odot}$, 13.5 pc, and -6.5 respectively \citep{Baumgardt19}. For clusters beyond 30 kpc with a measured central surface brightness, the average is 21 V magnitudes per square arcsecond \citep{Harris10}. These clusters provide an indication of what the properties of the undiscovered clusters could be. For a galactocentric distance of 145 kpc such a cluster would have an apparent size and apparent magnitude of 20.3$''$ and 14.2 respectively, assuming the galactic radius at the solar circle is 8 kpc. However bright cluster stars, like horizontal branch stars, might be difficult to resolve with an apparent magnitude of 21.5 \citep{deBoer99}.  For a more distant cluster that is 302 kpc from the Galactic centre, its  apparent size and apparent magnitude would be 9.0 $''$ and 16 respectively. Horizontal branch stars would have an apparent magnitude of 23.1. It should be noted that these are just estimates of the size and brightness of the undiscovered clusters based on the properties of known clusters with $r>30$ kpc. Given their large galactocentric distances, the undiscovered clusters are more likely to be lower in mass and more diffuse than known clusters in the Galaxy's halo. 

Since the location of these undiscovered clusters are unknown, they can only be discovered via large-scale surveys. The Sloan Digital Sky Survey \citep{Blanton17}, for example, has a magnitude limit in $V$ of 22.8 and a resolution on the order of  $0.7 ''/pixel$ \citep{Fukugita96,Gunn98,Doi10}. Hence the clusters would be detected by SDSS if the fell within the survey's footprint, but may be difficult to recognize as clusters. Future surveys with the Vera C. Rubin Observatory \citep{verarubin} and Euclid \footnote{https://www.euclid-ec.org/} will have a better chance of finding undiscovered Galactic Globular clusters with their lower magnitude limits and higher resolution. 

Whether or not these undiscovered clusters exist has important implications for the Milky Way. Additional outer clusters, which will almost certainly be extremely metal poor \citep{Youakim20}, will increase estimates of the number of star cluster forming satellites that initially formed around the Milky. One or more non-detections will indicate a departure from the outer Galactic globular cluster number density profile being a power-law. Such a departure may suggest outer clusters have been tidally stripped from the Milky Way or never formed in the first place. In either case, a complete outer cluster population will help constrain how the outer Galactic halo was assembled and mass estimates of the Milky Way's dark matter halo at large galactocentric radii.

\section*{Acknowledgements}
JW would like to thank Jo Bovy for helpful discussions regarding the manuscript.

\section*{Data Availability}
The data underlying this article are publicly available through the catalogues of \citet{Harris10}, \citet{Baumgardt19}, and \citet{Vasiliev19}.




\bibliographystyle{mnras}
\bibliography{ref2} 

\begin{thebibliography}{}
\makeatletter
\relax
\def\mn@urlcharsother{\let\do\@makeother \do\$\do\&\do\#\do\^\do\_\do\%\do\~}
\def\mn@doi{\begingroup\mn@urlcharsother \@ifnextchar [ {\mn@doi@}
  {\mn@doi@[]}}
\def\mn@doi@[#1]#2{\def\@tempa{#1}\ifx\@tempa\@empty \href
  {http://dx.doi.org/#2} {doi:#2}\else \href {http://dx.doi.org/#2} {#1}\fi
  \endgroup}
\def\mn@eprint#1#2{\mn@eprint@#1:#2::\@nil}
\def\mn@eprint@arXiv#1{\href {http://arxiv.org/abs/#1} {{\tt arXiv:#1}}}
\def\mn@eprint@dblp#1{\href {http://dblp.uni-trier.de/rec/bibtex/#1.xml}
  {dblp:#1}}
\def\mn@eprint@#1:#2:#3:#4\@nil{\def\@tempa {#1}\def\@tempb {#2}\def\@tempc
  {#3}\ifx \@tempc \@empty \let \@tempc \@tempb \let \@tempb \@tempa \fi \ifx
  \@tempb \@empty \def\@tempb {arXiv}\fi \@ifundefined
  {mn@eprint@\@tempb}{\@tempb:\@tempc}{\expandafter \expandafter \csname
  mn@eprint@\@tempb\endcsname \expandafter{\@tempc}}}

\bibitem[\protect\citeauthoryear{{Arp} \& {van den Bergh}}{{Arp} \& {van den
  Bergh}}{1960}]{Arp60}
{Arp} H.,  {van den Bergh} S.,  1960, \mn@doi [\pasp] {10.1086/127473}, \href
  {https://ui.adsabs.harvard.edu/abs/1960PASP...72...48A} {72, 48}

\bibitem[\protect\citeauthoryear{{Balbinot} et~al.,}{{Balbinot}
  et~al.}{2013}]{Balbinot2013}
{Balbinot} E.,  et~al., 2013, \mn@doi [\apj] {10.1088/0004-637X/767/2/101},
  \href {https://ui.adsabs.harvard.edu/abs/2013ApJ...767..101B} {767, 101}

\bibitem[\protect\citeauthoryear{{Baumgardt} \& {Makino}}{{Baumgardt} \&
  {Makino}}{2003}]{Baumgardt03}
{Baumgardt} H.,  {Makino} J.,  2003, \mn@doi [\mnras]
  {10.1046/j.1365-8711.2003.06286.x}, \href
  {https://ui.adsabs.harvard.edu/abs/2003MNRAS.340..227B} {340, 227}

\bibitem[\protect\citeauthoryear{{Baumgardt}, {Hilker}, {Sollima}  \&
  {Bellini}}{{Baumgardt} et~al.}{2019}]{Baumgardt19}
{Baumgardt} H.,  {Hilker} M.,  {Sollima} A.,   {Bellini} A.,  2019, \mn@doi
  [\mnras] {10.1093/mnras/sty2997}, \href
  {https://ui.adsabs.harvard.edu/abs/2019MNRAS.482.5138B} {482, 5138}

\bibitem[\protect\citeauthoryear{{Bechtol} et~al.,}{{Bechtol}
  et~al.}{2015}]{Bechtol2015_eri11}
{Bechtol} K.,  et~al., 2015, \mn@doi [\apj] {10.1088/0004-637X/807/1/50}, \href
  {https://ui.adsabs.harvard.edu/abs/2015ApJ...807...50B} {807, 50}

\bibitem[\protect\citeauthoryear{{Belokurov} et~al.,}{{Belokurov}
  et~al.}{2010}]{Belokurov2010}
{Belokurov} V.,  et~al., 2010, \mn@doi [\apjl] {10.1088/2041-8205/712/1/L103},
  \href {https://ui.adsabs.harvard.edu/abs/2010ApJ...712L.103B} {712, L103}

\bibitem[\protect\citeauthoryear{{Belokurov}, {Irwin}, {Koposov}, {Evans},
  {Gonzalez-Solares}, {Metcalfe}  \& {Shanks}}{{Belokurov}
  et~al.}{2014}]{Belokurov14}
{Belokurov} V.,  {Irwin} M.~J.,  {Koposov} S.~E.,  {Evans} N.~W.,
  {Gonzalez-Solares} E.,  {Metcalfe} N.,   {Shanks} T.,  2014, \mn@doi [\mnras]
  {10.1093/mnras/stu626}, \href
  {https://ui.adsabs.harvard.edu/abs/2014MNRAS.441.2124B} {441, 2124}

\bibitem[\protect\citeauthoryear{{Blanton} et~al.,}{{Blanton}
  et~al.}{2017}]{Blanton17}
{Blanton} M.~R.,  et~al., 2017, \mn@doi [\aj] {10.3847/1538-3881/aa7567}, \href
  {https://ui.adsabs.harvard.edu/abs/2017AJ....154...28B} {154, 28}

\bibitem[\protect\citeauthoryear{{Bovy}}{{Bovy}}{2015}]{Bovy15}
{Bovy} J.,  2015, \mn@doi [\apjs] {10.1088/0067-0049/216/2/29}, \href
  {https://ui.adsabs.harvard.edu/abs/2015ApJS..216...29B} {216, 29}

\bibitem[\protect\citeauthoryear{{Chambers} et~al.,}{{Chambers}
  et~al.}{2016}]{Chambers16}
{Chambers} K.~C.,  et~al., 2016, arXiv e-prints, \href
  {https://ui.adsabs.harvard.edu/abs/2016arXiv161205560C} {p. arXiv:1612.05560}

\bibitem[\protect\citeauthoryear{{Conn}, {Jerjen}, {Kim}  \& {Schirmer}}{{Conn}
  et~al.}{2018}]{Conn2018}
{Conn} B.~C.,  {Jerjen} H.,  {Kim} D.,   {Schirmer} M.,  2018, \mn@doi [\apj]
  {10.3847/1538-4357/aa9eda}, \href
  {https://ui.adsabs.harvard.edu/abs/2018ApJ...852...68C} {852, 68}

\bibitem[\protect\citeauthoryear{{Contenta}, {Gieles}, {Balbinot}  \&
  {Collins}}{{Contenta} et~al.}{2017a}]{Contenta17}
{Contenta} F.,  {Gieles} M.,  {Balbinot} E.,   {Collins} M. L.~M.,  2017a,
  \mn@doi [\mnras] {10.1093/mnras/stw3178}, \href
  {https://ui.adsabs.harvard.edu/abs/2017MNRAS.466.1741C} {466, 1741}

\bibitem[\protect\citeauthoryear{{Contenta}, {Gieles}, {Balbinot}  \&
  {Collins}}{{Contenta} et~al.}{2017b}]{Contenta2017}
{Contenta} F.,  {Gieles} M.,  {Balbinot} E.,   {Collins} M. L.~M.,  2017b,
  \mn@doi [\mnras] {10.1093/mnras/stw3178}, \href
  {https://ui.adsabs.harvard.edu/abs/2017MNRAS.466.1741C} {466, 1741}

\bibitem[\protect\citeauthoryear{{Davoust}, {Sharina}  \& {Donzelli}}{{Davoust}
  et~al.}{2011}]{Davoust2011}
{Davoust} E.,  {Sharina} M.~E.,   {Donzelli} C.~J.,  2011, \mn@doi [\aap]
  {10.1051/0004-6361/201015514}, \href
  {https://ui.adsabs.harvard.edu/abs/2011A&A...528A..70D} {528, A70}

\bibitem[\protect\citeauthoryear{{Doi} et~al.,}{{Doi} et~al.}{2010}]{Doi10}
{Doi} M.,  et~al., 2010, \mn@doi [\aj] {10.1088/0004-6256/139/4/1628}, \href
  {https://ui.adsabs.harvard.edu/abs/2010AJ....139.1628D} {139, 1628}

\bibitem[\protect\citeauthoryear{{Fukugita}, {Ichikawa}, {Gunn}, {Doi},
  {Shimasaku}  \& {Schneider}}{{Fukugita} et~al.}{1996}]{Fukugita96}
{Fukugita} M.,  {Ichikawa} T.,  {Gunn} J.~E.,  {Doi} M.,  {Shimasaku} K.,
  {Schneider} D.~P.,  1996, \mn@doi [\aj] {10.1086/117915}, \href
  {https://ui.adsabs.harvard.edu/abs/1996AJ....111.1748F} {111, 1748}

\bibitem[\protect\citeauthoryear{{Gaia Collaboration} et~al.,}{{Gaia
  Collaboration} et~al.}{2016}]{Gaia16a}
{Gaia Collaboration} et~al., 2016, \mn@doi [\aap]
  {10.1051/0004-6361/201629272}, \href
  {http://adsabs.harvard.edu/abs/2016A%26A...595A...1G} {595, A1}

\bibitem[\protect\citeauthoryear{{Gaia Collaboration} et~al.,}{{Gaia
  Collaboration} et~al.}{2018a}]{Gaia18}
{Gaia Collaboration} et~al., 2018a, \mn@doi [\aap]
  {10.1051/0004-6361/201833051}, \href
  {http://adsabs.harvard.edu/abs/2018A%26A...616A...1G} {616, A1}

\bibitem[\protect\citeauthoryear{{Gaia Collaboration} et~al.,}{{Gaia
  Collaboration} et~al.}{2018b}]{Gaia2018}
{Gaia Collaboration} et~al., 2018b, \mn@doi [\aap]
  {10.1051/0004-6361/201833051}, \href
  {https://ui.adsabs.harvard.edu/abs/2018A%26A...616A...1G} {616, A1}

\bibitem[\protect\citeauthoryear{{Gran} et~al.,}{{Gran}
  et~al.}{2019}]{Gran2019}
{Gran} F.,  et~al., 2019, \mn@doi [\aap] {10.1051/0004-6361/201834986}, \href
  {https://ui.adsabs.harvard.edu/abs/2019A&A...628A..45G} {628, A45}

\bibitem[\protect\citeauthoryear{{Gunn} et~al.,}{{Gunn} et~al.}{1998}]{Gunn98}
{Gunn} J.~E.,  et~al., 1998, \mn@doi [\aj] {10.1086/300645}, \href
  {https://ui.adsabs.harvard.edu/abs/1998AJ....116.3040G} {116, 3040}

\bibitem[\protect\citeauthoryear{{Halley}}{{Halley}}{1716}]{Halley1716}
{Halley} E.,  1716, Philosophical Transactions, 29, 391

\bibitem[\protect\citeauthoryear{{Harris}}{{Harris}}{1996}]{Harris96}
{Harris} W.~E.,  1996, \mn@doi [\aj] {10.1086/118116}, \href
  {https://ui.adsabs.harvard.edu/abs/1996AJ....112.1487H} {112, 1487}

\bibitem[\protect\citeauthoryear{{Harris}}{{Harris}}{2010}]{Harris10}
{Harris} W.~E.,  2010, arXiv e-prints, \href
  {https://ui.adsabs.harvard.edu/abs/2010arXiv1012.3224H} {p. arXiv:1012.3224}

\bibitem[\protect\citeauthoryear{{Herschel}}{{Herschel}}{1789}]{Herschel1789}
{Herschel} W.,  1789, Philosophical Transactions of the Royal Society of London
  Series I, \href {https://ui.adsabs.harvard.edu/abs/1789RSPT...79..212H} {79,
  212}

\bibitem[\protect\citeauthoryear{{Ibata}, {Gilmore}  \& {Irwin}}{{Ibata}
  et~al.}{1994}]{Ibata1994}
{Ibata} R.~A.,  {Gilmore} G.,   {Irwin} M.~J.,  1994, \mn@doi [\nat]
  {10.1038/370194a0}, \href
  {https://ui.adsabs.harvard.edu/abs/1994Natur.370..194I} {370, 194}

\bibitem[\protect\citeauthoryear{{Irrgang}, {Wilcox}, {Tucker}  \&
  {Schiefelbein}}{{Irrgang} et~al.}{2013}]{Irrgang13}
{Irrgang} A.,  {Wilcox} B.,  {Tucker} E.,   {Schiefelbein} L.,  2013, \mn@doi
  [\aap] {10.1051/0004-6361/201220540}, \href
  {https://ui.adsabs.harvard.edu/abs/2013A&A...549A.137I} {549, A137}

\bibitem[\protect\citeauthoryear{{Ivezi{\'c}} et~al.,}{{Ivezi{\'c}}
  et~al.}{2019}]{verarubin}
{Ivezi{\'c}} {\v Z}.,  et~al., 2019, \mn@doi [\apj] {10.3847/1538-4357/ab042c},
  \href {http://adsabs.harvard.edu/abs/2019ApJ...873..111I} {873, 111}

\bibitem[\protect\citeauthoryear{{Kim} \& {Jerjen}}{{Kim} \&
  {Jerjen}}{2015}]{Kim2015}
{Kim} D.,  {Jerjen} H.,  2015, \mn@doi [\apj] {10.1088/0004-637X/799/1/73},
  \href {https://ui.adsabs.harvard.edu/abs/2015ApJ...799...73K} {799, 73}

\bibitem[\protect\citeauthoryear{{Kim}, {Jerjen}, {Milone}, {Mackey}  \& {Da
  Costa}}{{Kim} et~al.}{2015}]{Kim2015b}
{Kim} D.,  {Jerjen} H.,  {Milone} A.~P.,  {Mackey} D.,   {Da Costa} G.~S.,
  2015, \mn@doi [\apj] {10.1088/0004-637X/803/2/63}, \href
  {https://ui.adsabs.harvard.edu/abs/2015ApJ...803...63K} {803, 63}

\bibitem[\protect\citeauthoryear{{Kim}, {Jerjen}, {Mackey}, {Da Costa}  \&
  {Milone}}{{Kim} et~al.}{2016}]{Kim2016}
{Kim} D.,  {Jerjen} H.,  {Mackey} D.,  {Da Costa} G.~S.,   {Milone} A.~P.,
  2016, \mn@doi [\apj] {10.3847/0004-637X/820/2/119}, \href
  {https://ui.adsabs.harvard.edu/abs/2016ApJ...820..119K} {820, 119}

\bibitem[\protect\citeauthoryear{{Kruijssen} \& {Mieske}}{{Kruijssen} \&
  {Mieske}}{2009}]{Kruijssen09}
{Kruijssen} J.~M.~D.,  {Mieske} S.,  2009, \mn@doi [\aap]
  {10.1051/0004-6361/200811453}, \href
  {https://ui.adsabs.harvard.edu/abs/2009A&A...500..785K} {500, 785}

\bibitem[\protect\citeauthoryear{{Kruijssen} et~al.,}{{Kruijssen}
  et~al.}{2020}]{Kruijssen20}
{Kruijssen} J.~M.~D.,  et~al., 2020, arXiv e-prints, \href
  {https://ui.adsabs.harvard.edu/abs/2020arXiv200301119K} {p. arXiv:2003.01119}

\bibitem[\protect\citeauthoryear{{Laevens} et~al.,}{{Laevens}
  et~al.}{2014}]{Laevens2014}
{Laevens} B. P.~M.,  et~al., 2014, \mn@doi [\apjl]
  {10.1088/2041-8205/786/1/L3}, \href
  {https://ui.adsabs.harvard.edu/abs/2014ApJ...786L...3L} {786, L3}

\bibitem[\protect\citeauthoryear{{Laevens} et~al.,}{{Laevens}
  et~al.}{2015}]{Laevens2015}
{Laevens} B. P.~M.,  et~al., 2015, \mn@doi [\apj] {10.1088/0004-637X/813/1/44},
  \href {https://ui.adsabs.harvard.edu/abs/2015ApJ...813...44L} {813, 44}

\bibitem[\protect\citeauthoryear{{Larsen}, {Romanowsky}, {Brodie}  \&
  {Wasserman}}{{Larsen} et~al.}{2020}]{Larsen2020}
{Larsen} S.~S.,  {Romanowsky} A.~J.,  {Brodie} J.~P.,   {Wasserman} A.,  2020,
  arXiv e-prints, \href {https://ui.adsabs.harvard.edu/abs/2020arXiv201007395L}
  {p. arXiv:2010.07395}

\bibitem[\protect\citeauthoryear{{Luque} et~al.,}{{Luque}
  et~al.}{2016}]{Luque2016_des1}
{Luque} E.,  et~al., 2016, \mn@doi [\mnras] {10.1093/mnras/stw302}, \href
  {https://ui.adsabs.harvard.edu/abs/2016MNRAS.458..603L} {458, 603}

\bibitem[\protect\citeauthoryear{{Lynn}}{{Lynn}}{1886}]{Lynn1886}
{Lynn} W.~T.,  1886, The Observatory, \href
  {https://ui.adsabs.harvard.edu/abs/1886Obs.....9..163L} {9, 163}

\bibitem[\protect\citeauthoryear{{Madore} \& {Arp}}{{Madore} \&
  {Arp}}{1979}]{Madore79}
{Madore} B.~F.,  {Arp} H.~C.,  1979, \mn@doi [\apjl] {10.1086/182876}, \href
  {https://ui.adsabs.harvard.edu/abs/1979ApJ...227L.103M} {227, L103}

\bibitem[\protect\citeauthoryear{{Massari}, {Koppelman}  \& {Helmi}}{{Massari}
  et~al.}{2019}]{Massari19}
{Massari} D.,  {Koppelman} H.~H.,   {Helmi} A.,  2019, \mn@doi [\aap]
  {10.1051/0004-6361/201936135}, \href
  {https://ui.adsabs.harvard.edu/abs/2019A&A...630L...4M} {630, L4}

\bibitem[\protect\citeauthoryear{{McMillan}}{{McMillan}}{2017}]{McMillan17}
{McMillan} P.~J.,  2017, \mn@doi [\mnras] {10.1093/mnras/stw2759}, \href
  {https://ui.adsabs.harvard.edu/abs/2017MNRAS.465...76M} {465, 76}

\bibitem[\protect\citeauthoryear{{McMillan} \& {Dehnen}}{{McMillan} \&
  {Dehnen}}{2007}]{mcmillan07}
{McMillan} P.~J.,  {Dehnen} W.,  2007, \mn@doi [\mnras]
  {10.1111/j.1365-2966.2007.11753.x}, \href
  {https://ui.adsabs.harvard.edu/abs/2007MNRAS.378..541M} {378, 541}

\bibitem[\protect\citeauthoryear{{Minniti} et~al.,}{{Minniti}
  et~al.}{2010}]{Minniti10}
{Minniti} D.,  et~al., 2010, \mn@doi [\na] {10.1016/j.newast.2009.12.002},
  \href {https://ui.adsabs.harvard.edu/abs/2010NewA...15..433M} {15, 433}

\bibitem[\protect\citeauthoryear{{Mu{\~n}oz}, {Geha}, {C{\^o}t{\'e}}, {Vargas},
  {Santana}, {Stetson}, {Simon}  \& {Djorgovski}}{{Mu{\~n}oz}
  et~al.}{2012}]{Munoz2012}
{Mu{\~n}oz} R.~R.,  {Geha} M.,  {C{\^o}t{\'e}} P.,  {Vargas} L.~C.,  {Santana}
  F.~A.,  {Stetson} P.,  {Simon} J.~D.,   {Djorgovski} S.~G.,  2012, \mn@doi
  [\apjl] {10.1088/2041-8205/753/1/L15}, \href
  {https://ui.adsabs.harvard.edu/abs/2012ApJ...753L..15M} {753, L15}

\bibitem[\protect\citeauthoryear{Myeong, Evans, Belokurov, Sanders  \&
  Koposov}{Myeong et~al.}{2018}]{Myeong18}
Myeong G.~C.,  Evans N.~W.,  Belokurov V.,  Sanders J.~L.,   Koposov S.~E.,
  2018, \mn@doi [The Astrophysical Journal] {10.3847/2041-8213/aad7f7}, 863,
  L28

\bibitem[\protect\citeauthoryear{{Plaskett}}{{Plaskett}}{1936}]{Plaskett1939}
{Plaskett} J.~S.,  1936, \jrasc, \href
  {https://ui.adsabs.harvard.edu/abs/1936JRASC..30..153P} {30, 153}

\bibitem[\protect\citeauthoryear{{Ryu} \& {Lee}}{{Ryu} \& {Lee}}{2018}]{Ryu18}
{Ryu} J.,  {Lee} M.~G.,  2018, \mn@doi [\apjl] {10.3847/2041-8213/aad8b7},
  \href {https://ui.adsabs.harvard.edu/abs/2018ApJ...863L..38R} {863, L38}

\bibitem[\protect\citeauthoryear{{Shapley}}{{Shapley}}{1918a}]{Shapley18a}
{Shapley} H.,  1918a, \mn@doi [\apj] {10.1086/142419}, \href
  {https://ui.adsabs.harvard.edu/abs/1918ApJ....48...89S} {48, 89}

\bibitem[\protect\citeauthoryear{{Shapley}}{{Shapley}}{1918b}]{Shapley18b}
{Shapley} H.,  1918b, \mn@doi [\apj] {10.1086/142423}, \href
  {https://ui.adsabs.harvard.edu/abs/1918ApJ....48..154S} {48, 154}

\bibitem[\protect\citeauthoryear{{Vasiliev}}{{Vasiliev}}{2019}]{Vasiliev19}
{Vasiliev} E.,  2019, \mn@doi [\mnras] {10.1093/mnras/stz171}, \href
  {https://ui.adsabs.harvard.edu/abs/2019MNRAS.484.2832V} {484, 2832}

\bibitem[\protect\citeauthoryear{{Youakim} et~al.,}{{Youakim}
  et~al.}{2020}]{Youakim20}
{Youakim} K.,  et~al., 2020, \mn@doi [\mnras] {10.1093/mnras/stz3619}, \href
  {https://ui.adsabs.harvard.edu/abs/2020MNRAS.492.4986Y} {492, 4986}

\bibitem[\protect\citeauthoryear{{de Boer}}{{de Boer}}{1999}]{deBoer99}
{de Boer} K.~S.,  1999, in {Egret} D.,  {Heck} A.,  eds,  Astronomical Society
  of the Pacific Conference Series Vol. 167, Harmonizing Cosmic Distance Scales
  in a Post-HIPPARCOS Era. pp 129--139 (\mn@eprint {arXiv} {astro-ph/9811077})

\makeatother
\end{thebibliography}


\bsp	
\label{lastpage}
\end{document}